\documentclass[structabstract]{aa}  

\usepackage{tabularx}
\usepackage{graphicx}
\usepackage{my_natbib}
\usepackage{txfonts}
\usepackage{longtable}
\usepackage{multirow}
\usepackage{lscape}

\newcommand{\hersc}{{\it Herschel}}
\newcommand{\lab}{LABOCA}

\newcommand{\spitz}{{\it  Spitzer}}

\newcommand{\iras}{{\it IRAS}}

\newcommand{\msun}{$M_\odot$}

\newcommand{\mic}{$\mu$m}

\renewcommand*{\figurename}{ Fig.}
\renewcommand*{\tablename}{Table}
\newlength{\pointwidth}
\def\revised{}
\def\revisedbis{}

\begin{document}

  \title{Probing the Dust Properties of Galaxies at Submillimetre Wavelengths }
  \subtitle{II. Dust-to-gas mass ratio trends with metallicity and the submm excess in dwarf galaxies}
  
  \author{Maud Galametz 
  	       \inst{1}~,
	       Suzanne C. Madden 
	       \inst{1}~,
	       Fr{\'e}d{\'e}ric Galliano
	       \inst{1}~,
	       Sacha Hony
	       \inst{1}~,
	       George J. Bendo
	       \inst{2}~,
	      Marc Sauvage
	       \inst{1}~}

	      \institute{Laboratoire AIM, CEA, Universit\'{e} Paris Diderot, IRFU/Service d'Astrophysique, Bat. 709, 91191 Gif-sur-Yvette, France,  \\
              \email{mgalamet@ast.cam.ac.uk} 
              \and
              Jodrell Bank Centre for Astrophysics, University of Manchester, Alan Turing Building, Manchester, M13 9PL, UK 
 }


 \abstract
 {}
 {We are studying the effects of submm observations on the total dust mass and thus dust-to-gas mass ratio measurements.}
 {We gather a wide sample of galaxies that have been observed at submillimeter (submm) wavelengths to model their Spectral Energy Distributions using submm observations ($>$160 \mic) and then without submm observational constraints in order to quantify the error on the dust mass when submm data are not available. Our model does not make strong assumptions on the dust temperature distribution to precisely avoid submm biaises in the study. Our sample includes 52 galaxies observed at submm wavelengths. Out of these, 9 galaxies show an excess in submm which is not accounted for in our fiducial model, most of these galaxies being low-metallicity dwarfs. We chose to add an independant very cold dust component (T=10K) to account for this excess.}
 {We find that metal-rich galaxies modelled with submm data often show lower dust masses than when modelled without submm data. Indeed, these galaxies usually have dust SEDs that peaks at longer wavelengths and require constraints above 160 \mic\ to correctly position the peak and sample the submillimeter slope of their SEDs and thus correctly cover the dust temperature distribution. On the other hand, some metal-poor dwarf galaxies modelled with submm data show higher dust masses than when modelled without submm data. Using submm constraints for the dust mass estimates, we find a tightened correlation of the dust-to-gas mass ratio with the metallicity of the galaxies. We also often find that when there is a submm excess present, it occurs preferentially in low-metallicity galaxies. We analyse the conditions for the presence of this excess and find a relation between the 160/850 \mic\ ratio and the submm excess.}
 {}
  

     \authorrunning{Galametz M. et al}
     \titlerunning{Submm excess and dust-to-gas mass ratio vs metallicity}

 \maketitle


\section{Introduction}

Interstellar Medium (ISM) dust is formed in the ejected elements of dying stars. The location of the actual formation of dust grains, whether in the ejecta or farther away in the ISM, is still debated. Within the ISM, dust undergoes constructive processes of accretion or coagulation \citep{Stepnik2003,Hirashita2009} or destructive/disruptive processes of vaporisation, fragmentation, erosion and shattering \citep{Jones1994,Jones1996}. Dust grains are, moreover, incorporated in the formation of new stars through gravitational collapse of gas and dust clouds. Dust grains also participate in the synthesis of molecules on their surface \citep[e.g.][]{Hasegawa1993, Vidali2004}, leading to the formation of giant molecular clouds within the galaxy. Dust finally plays a key role in the overall emission of a galaxy since it absorbs stellar and ionized gas radiation to re-emit it from Infrared (IR) to submillimeter (submm) wavelengths, thus shaping its overall Spectral Energy Distibution (SED). Dust thus enables us to have access to the most embedded star forming regions of a galaxy which makes it an indirect but reliable tracer of its star formation. Dust grains are thus, at the same time, relics as well as engines of the evolution of the galaxy.

The ISO and \spitz\ telescopes have, in the last few years, opened up the view of the interstellar dust in mid-infrared (mid-IR) to far-infrared (far-IR) wavelengths. They enabled us to improve our knowledge of the dust properties of the ISM such as the distribution of the different dust components (Polycyclic Aromatic Hydrocarbons (PAHs), silicate, graphite) and how they are affected by metallicity ~\citep{Galliano_Dwek_Chanial_2008, Engelbracht2008, Bendo2008, Munoz2009_2}. 

Many studies have also been performed to investigate the relation between the Dust-to-Gas mass ratio (D/G) and metal enrichment of galaxies, the D/G of a galaxy being the main output of dust evolution models and a crucial key to understand the cycle of matter within galaxies. Early studies suggested that the D/G should be proportional to the metallicity of the galaxy  \citep{Franco_Cox_1986}. Investigations in the Milky Way (MW) found a D/G of $\sim$ 10$^{-2}$ for our galaxy \citep{Sodroski1997}. Further analysis into depletion of metals from the ISM or into comparisons of dust extinction versus hydrogen column densities have both shown that the D/G of the MW falls within the range of 1-5 $\times$ 10$^{-2}$ \citep[see][for more complete details]{Whittet2003,Kruegel2003,Kruegel2008}. To investigate the relation between the metallicity and the D/G, \citet{Lisenfeld_Ferrara_1998} studied a broad sample of dwarf irregular (dIrrs) and blue compact dwarfs (BCDs) from {\it Infrared Astronomical Satellite} (\iras) observations at 60 and 100 \mic. A logarithmic correlation between D/G and the oxygen abundance was found for the dIrrs: 12 + log (O/H) $\propto$ (0.52$\pm$0.25) $\times$ log (M$_{dust}$/M$_{HI}$), but no relation was found for the BCDs which are more actively forming stars. Therefore, they observed that metals are less effectively incorporated into dust in dwarf galaxies than in more metal-rich galaxies. 
Using the Spitzer Infrared Nearby Galaxies Survey sample (SINGS), \citet{Engelbracht2008} also found a steep decrease of the D/G from solar metallicity to a metallicity of 8 and note that the D/G seems to flatten out at low metallicity. With the same sample, \citet{Munoz2009_2} showed that the D/G decreases with galactocentric radius. \citet{Bendo2010_NGC2403} also compare radial variations in the D/G with the oxygen abundance gradients and found a good correspondence, although the D/G radial variations seems to be steeper. Recent studies in the low-metallicity Magellanic Clouds have shown that their D/G are lower than that estimated for more metal-rich environments \citep{Bernard2008,Gordon2009}, consistent with what should be expected from the previous relation between the D/G and the metallicity. 

\citet{James2002}, using JCMT/SCUBA observations at 850 \mic\ observations, concluded that the fraction of metals incorporated into dust was a universal constant, that is to say similar for dwarf and giant galaxies, which was in contradiction to the \citet{Lisenfeld_Ferrara_1998} conclusions. They also noted that their D/G mass ratios estimated with submm observations are usually higher than those previously estimated by \citet{Lisenfeld_Ferrara_1998} for the same galaxies. \citet{Galliano2003, Galliano2005} also showed that taking the very cold dust traced in some low metallicity galaxies by SCUBA or MAMBO, the dust-to-metals ratios are closer to the Galactic value. \citet{Galametz2009} present the SED models of 4 low-metallicity galaxies observed with the APEX/\lab\ instrument at 870 \mic\ and noticed that the dust masses estimated from their models increase when using the submm constraints. On the contrary, \citet{Draine2007} found that the dust masses of the subset of their SINGS sample observed with SCUBA,  which are metal-rich galaxies, are rather unaffected by submm constraints and that omitting the submm data only moderately increases the D/G. Nevertheless, \citet{Vlahakis2005} used the SCUBA Local Universe Galaxy Survey (SLUGS) to confirm that most of the dust mass of their galaxies was contained in cold grains. \citet{Willmer2009} completed the coverage of the far-IR/submm SED of the SLUGS sample with \spitz\ data and found that these galaxies show flatter $\nu$F$_{\nu}$(160 \mic)/$\nu$F$_{\nu}$(850 \mic) slopes than the SINGS sample in which galaxies with large amounts of cold dust seem to be under-represented. \\

Many issues thus still remain open to correctly characterize the D/G with metallicity, among which is the basic question of the accurate quantification of the total dust mass of a galaxy and the real impact of submm constraints in the assessment of this mass. To complicate the mass determination issue, there have been findings of excess emission in the submm wavelength regime \citep[][among others]{Galliano2003,Dumke2004, Galliano2005,Bendo2006,Marleau2006, Galametz2009, Grossi2010, OHalloran2010, Israel2010, Bot2010}. 

{\revisedbis Several hypotheses have be studied to explain this excess by a variation of the dust emissivity in that submm regime:

Models of \citet{Meny2007} and observations of \citet{Dupac2003} found an effective decrease in the submm emissivity index as the dust temperature increases.  \citet{Shetty2009}, however, express caution in the inverse temperature - $\beta$ interpretation, showing that flux uncertainties, especially in the Rayleigh-Jeans regime, can affect the results for the SED fits as far as temperature and emissivity are concerned. It is still not clear what would be the nature of this new grain population or which processes could lead to their predominance in dwarf galaxies.
Recently, \citet{Paradis2009} interpreted the increase of the far-IR emissivity in molecular clouds of the Milky Way containing cold dust by fractal aggregates of individual amorphous grains. They note that an increase of a factor of 3-4 in the dust emissivity is required to explain the unusually low dust temperatures observed by dust aggregation. This explanation of a submm excess by grain coagulation was already suggested by \citet{Bazell1990} (theorethical approach) and \citet{Stepnik2001} (using the baloon-borne experiment ProNaos). We note that this study is limited to solar metallicity environments for now.
\citet{Serra_Dias_Cano2008} also note that graphite can not reproduce the carbonaceous grain sputtering in interstellar shock waves and cautionned the use of graphite in dust models. They proposed that hydrogenated amorphous carbon could be the most probable form of carbonaceous grain material. Amorphous carbons have a lower emissivity index which could be responsible for a flattening of the SED at submm wavelength compared to current SED models using graphite \citep[i.e.][]{Meixner2010,Galametz2010}. This scenario alone would nevertheless be unable to reproduce the knee observed in the submm regime of the SEDs of some low-metallicity galaxies observed at 450 and 850 \mic\ such as II~Zw~40 (see Fig.~\ref{SED_CD}).

Other explanations try to explain the excess usually observed beyond 400 \mic\ by a different population of grains: 

\citet{Lisenfeld2001} and studies of \citet{Zhu2009} on extreme extragalactic environments suggested that the submm excess could originate from an enhanced very small grain abundance with an emissivity index $\beta$=1. 
 \citet{Ferrara1994} showed that `spinning dust' located in the ionized gas of many galaxies were producing radio emission. Studies by \citet{Anderson1993}, \citet{Draine_Lazarian_1998}, \citet{Hoang2010}, \citet{Silsbee2010} or the recent {\it Planck} results \citep{Planck_collabo_2011_SpinningDust} aimed to characterise this component which explains most of the 10-100 GHz emission, also called the anomalous foreground. \citet{Ysard2010} recently suggested that their peak frequency could depend on different parameters (radiation field intensity, size distribution of the dust species, dipole moment distribution, physical parameters of the gas phases etc.) and thus that spinning dust emission could be responsible for the excess at submm wavelengths observed in some galaxies, as recently suggested by \citet{Bot2010} for the Large Magellanic Cloud. Spinning dust emission being generated by very rapidly spinning and assymetric small dust grains, PAHs are often proposed as possible carriers. However, this assumption seems to be in contradiction with the lack of PAH features observed in low-metallicity galaxies in which submm excess is usually detected. 

This excess could finally be interpreted as a very cold ($\le$10K) dust component \citep[e.g.][]{Galliano2003, Galliano2005,Galametz2009}. This latter hypothesis could explain the break observed in the submm regime of some SEDs, especially in low-metallicity galaxies. However, its use in SED models usually implies a significant increase of the total dust mass, with more than 50$\%$ of the total dust mass residing in this very cold component, leading to unrealistic D/G in some galaxies \citep[e.g.][]{Dumke2004,Bendo2006_Sombrero}, if we are taking into account the total gas reservoir detected.
 \\

The ISO and \spitz\ telescopes offered a vision of the IR emission limited to 160 \mic\ and thus do not place constraints on the bulk of the coldest ($<$15K) dust of the galaxies, that could represent a large amount of the dust mass of galaxies. For many galaxies, the \hersc\ telescope, more specifically the SPIRE instrument observing from 250 to 500 \mic, will better constrain the submm slope and thus help to disentangle between those scenarios. The issues discussed here will continue to be of concern throughout the \hersc, {\it Planck} and {\it ALMA} era, especially for low-metallicity objects not detected. Thus, this study lays the ground work for an eventual broader study from which these future data can build on. \\
}

We perform a systematic study of the dust SEDs of galaxies spanning of a wide range of metallicities with the following objectives:

1) compare the estimates of the total dust mass of the galaxies when the SED modelling has submm constraints, to the dust masses determined when the submm observations are not used (as if they are not available).

2) using only the sample modeled with submm data, to have a reliable dust mass, determine the D/G as a function of metallicity and compare with chemical evolution models.

3) identify the galaxies of the sample which show a submm excess.

This paper is organised as follows: in $\S$ 2, we describe the preliminary study we perform on the D/G as a function of metallicity by gathering dust and gas masses from the literature for a large sample of galaxies. In $\S$ 3, we describe the common SED modelling technique we apply to remodel a subsample of these galaxies in a unique way with and without submm constraints. Finally, in $\S$ 4, we analyse the differences obtained in the dust masses found with or without submm constraints and analyse the conditions under which a submm excess is found in galaxies and relate this to the metallicity. We finally summarize the main conclusions of the study in $\S$ 5.

\section{Genesis of our study}

For this $\S$, we perform a preliminary study based on data directly taken from the literature, in particular, the dust masses.   

\subsection{The sample}

We have gathered published dust and gas masses of a broad sample of galaxies, some of them observed with submm telescopes, to study the relation of D/G with metallicity. The dust and gas masses used in this preliminary study are published in:

\begin{description}

\item  \citet{Lisenfeld_Ferrara_1998}: This sample groups together dwarf irregular galaxies (dIrrs) and BCDs. Their SEDs were performed using only \iras\ data and a single temperature, modified black body model. Their dust mass estimates are thus considered as lower limits to the total dust masses of the galaxies. 
\item \citet{James2002}: In this sample, the submm observations were performed with JCMT/SCUBA at 850 \mic. They fitted their data with the \citet{Dunne_Eales_2001} two-component modified blackbodies to derive the total dust mass of their galaxies.
\item \citet{Hirashita2008}: This sample of eight low-metallicity galaxies was observed with {\it AKARI} at 65, 90, 140 and 160 \mic. They derived a total mass of dust for their galaxies from the far-IR emission using a single temperature modified blackbody model, using the mass absorption coefficient of dust grains taken from \citet{Hildebrand1983}.
\item \citet{Draine2007}: Their sample \citep[SINGS, see][]{Kennicutt2003} includes spirals, ellipticals, starburst galaxies and both metal-rich and metal-poor galaxies. Their dust SED modelling use the \citet{Li_Draine_2001} dust properties and is performed with \spitz\ IRAC and MIPS data, with some of their galaxies observed with SCUBA at 450 or/and 850 \mic. 
\item \citet{Engelbracht2008}: Their galaxies, observed with \spitz, cover a wide range of metallicity values (from 12+log(O/H)=7.1 for SBS 0335-052 to 8.85 for IC342). Their dust masses are estimated using the standard formula and the absorption coefficients from \citet{Li_Draine_2001}.
\item \citet{Galliano_Dwek_Chanial_2008}: A subset of their sample had been observed with SCUBA 450 \mic\ and/or 850 \mic. All of their SED models included either the \spitz/IRS spectra or the ISOCAM spectra for optimum contraints in the mid-IR. Their model includes a detailed decomposition of dust emission into its gas-phase components.
\item \citet{Galametz2009}: their sample contains 3 low-metalllicity sources observed with the \spitz\ telescope and \lab, a 870 \mic\ multi-channel bolometer array available on the APEX telescope in Chile.

\end{description}

In these papers, the gas masses are estimated using H{\sc i} 21cm line measurements and the H$_2$ mass determined from CO observations, when available.

\subsection{The relation between the dust-to-gas ratio and metallicity}

Figure~\ref{Dust-to-gas_Metallicity_1} shows the D/G of these galaxies as a function of metallicity. We overlay several chemical evolution models on this figure: the one-zone single-phase chemical evolution model of \citet{Dwek1998} presented in \citet{Galliano_Dwek_Chanial_2008} and the dust formation models of  \citet{Edmunds2001}.  \citet{Edmunds2001} note that they adopt a less detailed quasi-analytic model compared to \citet{Dwek1998}. Their models include mantle growth in the ISM  and take into account the grain core production in both supernovae and giant stars. These two modes of dust formation are plotted in Fig.~\ref{Dust-to-gas_Metallicity_1}. The solid line corresponds to dust produced in low and intermediate mass stars and the dotted line corresponds to dust production strongly affected by the role of supernovae.

The filled symbols localise the galaxies for which dust masses are estimated using submm constraints. These galaxies tend to follow the \citet{Edmunds2001} and \citet{Dwek1998} models and clearly differ from the linear regression performed with the whole sample (dashed line), i.e. which also include galaxies that were not observed at submm wavelengths. We particularly noticed that low-metallicity galaxies (12+log(O/H)$<$8.5) whose dust masses were estimated with submm constraints (filled symbols) have higher D/G than galaxies modelled without submm constraints (empty symbols, crosses or asterisks). This could imply that for dwarf galaxies, the use of submm constraints leads to higher dust mass predictions. We further investigate this behaviour in this study. \\

\subsection{The limitations of the samples}

Of course, these various samples do not necessarily have any selection criteria in common. Some of the surveys were deliberately targeting dwarf galaxies yet others are covering a wider variety of galaxies. For instance, the SINGS sample is designed with a broad range of galaxies evenly filling a three-parameter space defined by morphology, infrared luminosity and L$_{FIR}$/L$_{optical}$. The dust models and SED fitting {\revisedbis procedure} used to derive D/G are different from one sample to another, which could lead to biases in the analysis and interpretation. Moreover, all of the galaxies used in this broad sample have not been observed at submm wavelengths, data that are necessary to put constraints on the coldest phases of the dust which could represent a large amount of the total dust mass. {\revisedbis Such a global study should quantify the total gas mass in a uniform manner. We discuss this issue in Section 4.2.1. }

\begin{figure*}
    \centering
    \begin{tabular}{ c }
        \includegraphics[width=12.5cm ,height=9cm]{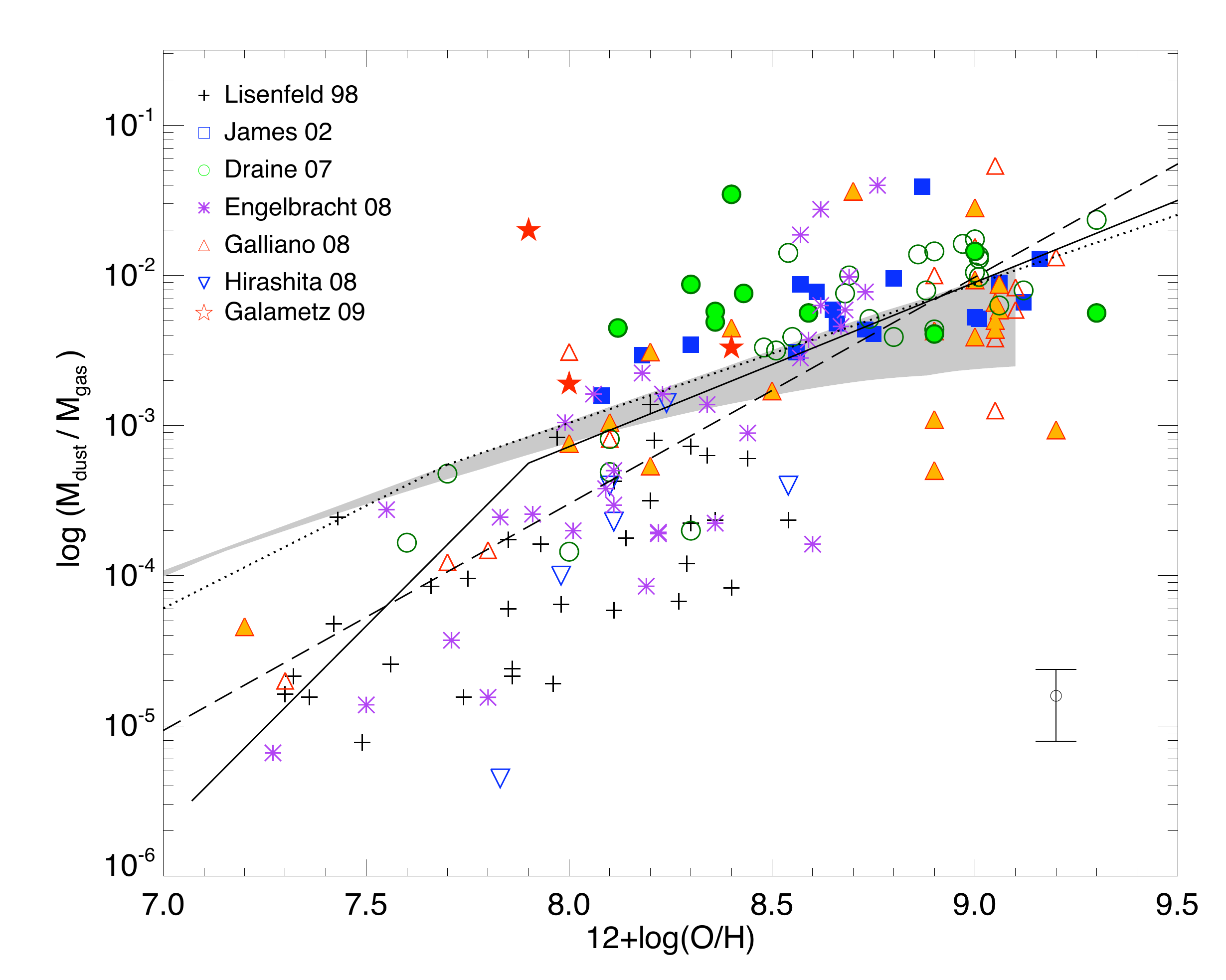} \\
         \end{tabular}
    \caption{Dust-to-gas mass ratio as a function of oxygen abundance. Black crosses show the \citet{Lisenfeld_Ferrara_1998} sample of Blue Compact Galaxies and dwarf Irregulars. Purple asterisks show the \citet{Engelbracht2008} sample using \spitz\ data. Blue squares indicate the \citet{James2002} galaxies. Green circles show the galaxies of the SINGS survey \citep{Draine2007} for which metallicity is already published. Orange upward triangles are the \citet{Galliano_Dwek_Chanial_2008} sample, blue downward triangles are the \citet{Hirashita2008} sample. Stars finally indicate the location of the three low-metallicity galaxies Haro~11, NGC~1705 and Mrk~1089 of ~\citet{Galametz2009}. Filled triangles, squares, circles or stars indicate when submm data were used in the SED modelling to estimate the total dust mass of the SED. The solid line and the dotted line are the predictions of the dust formation models from \citet{Edmunds2001}: the solid line shows a model in which dust is produced in low and intermediate mass stars while the dotted line shows the model in which supernovae play an important role in the production of dust. The gray stripe is the expectation of the one-zone, single-phase chemical evolution model of \citet{Dwek1998} presented in \citet{Galliano_Dwek_Chanial_2008}. The dashed line shows the linear regression of the whole sample. The D/G errors can be significant for many galaxies but should not exceed $\pm$ 50$\%$. (the $\pm$ 50$\%$ error bars are plotted on the bottom right part of the plot)}
    \label{Dust-to-gas_Metallicity_1}
\end{figure*}

In conclusion, analyzing the dust and gas masses of these galaxies directly given by the literature as one broad collection should be limited. We decided {\it 1)} to reprocess this analysis using a common realistic SED model to avoid biases in comparing galaxies analysed with different SED models; {\it 2)} to restrict our study to galaxies observed at submm wavelengths to determine, for a set of fixed dust properties (emissivity, composition), the influence of submm constraints on the estimates of the total dust mass of the galaxies. Thus we use a subsample of the galaxies from Fig.~\ref{Dust-to-gas_Metallicity_1} in the following study, namely galaxies which have submm observations.

\section{The SED models}

\subsection {The wavelength coverage}

From the previous full sample described in $\S$ 2, we only keep galaxies for which submm data are available, namely observational constraints beyond 200 \mic. The restricted sample is presented in Table~\ref{Effects_of_the_submm_Table} and gathers subsamples of the \citet{James2002}, \citet{Draine2007}, \citet{Galliano_Dwek_Chanial_2008} and \citet{Galametz2009} samples.

Metallicities, distances and submm 450 and/or 850 (or 870) \mic\ observations for this subsample are summarized in Table~\ref{Effects_of_the_submm_Table}. 

\subsubsection {Submm flux densities}

Most of the submm data we used are flux densities at 450 and/or 850 \mic\ from SCUBA or 870 \mic\ from LABOCA. A few galaxies of our sample were not observed with SCUBA or LABOCA but with other submm instruments. M83 was observed at 540 \mic\ with the 4-m telescope at Cerro Tololo Inter American Observatory \citep{Hildebrand1977}. NGC~1068 was observed at 350 \mic\ and 390 \mic, respectively with the Submm High Angular Resolution Camera (SHARC) \citep{Benford1999} and with the University of Hawaii 2.2-m telescope at the Mauna Kea Observatory \citep{Hildebrand1977}. NGC~891 was observed at 350 and 1300 \mic, with the 3-m Infrared Telescope Facility (IRTF) and the 88" telescope of the University of Hawaii on Mauna Kea \citep{Chini1986}. Finally, IC~342 was observed at 1mm with the Palomar 5-m telescope \citep{Elias1978}. Only an upper limit was derived for this galaxy.

We would also like to warn the reader about the quality of the submm fluxes derived for some of the SINGS galaxies used in this paper. Indeed, \citet{Draine2007} warned that the galaxy NGC~6946 was scan mapped with SCUBA and that the data processing could have removed some extended emission for this galaxy, leading to an underestimate of submm fluxes. Furthermore, in the case of NGC~1097, SCUBA observations only cover the galaxy center and the total submm fluxes were estimated by inferring how much flux resides in the outer regions not observed, assuming a constant MIPS-to-submm ratio \citep{Dale2007}. This assumption could be incorrect if radial color variations are present. SCUBA maps of NGC~3521, NGC~3627, NGC~4536 and NGC~7331 also have small field of views. The submm fluxes presented in \citet{Dale2007} should therefore be used with caution in a global analysis of the galaxies. 

\subsubsection {Optical and infrared flux densities}

{\it 2MASS} data constrain the stellar contribution of the SEDs. {\it 2MASS} flux densities are taken from the 2MASS Large Galaxy Atlas \citep{Jarrett2003},  \citet{Galametz2009} and \citet{Dale2007}. \spitz/IRAC and MIPS flux densities are used to sample the infrared part of the SED, when available. The \spitz\ flux densities used in the following study are also given in \citet{Dale2007}, \citet{Engelbracht2008} and \citet{Galametz2009}. {\revisedbis For the SINGS galaxies used in this paper and for which far-IR data are critical when submm data are not available, the \spitz\ data enable us to  sample both the peak position and the beginning of the submm slope.}
For the \citet{Galliano_Dwek_Chanial_2008} sample, \spitz/IRS spectral information is, in most cases, available and used in our modelling. These spectra are presented in \citet{Galliano_Dwek_Chanial_2008}.  

We use the best available \iras\ fluxes. The \iras\ flux densities of the \citet{Draine2007} galaxies were obtained from SCANPI and the \iras\ High Resolution Image Restoration Atlas \citep[HIRES;][]{Surace2004}. Those of the \citet{James2002} sample are coming from the \iras\ Revised Bright Galaxy Sample \citep{Sanders2003}, HIRES for the galaxies NGC~3994 and NGC~3995 and from the \iras\ Faint Source Catalog  \citep{Moshir1990} for NGC~4670. The \iras\ flux densities of \citet{Galliano_Dwek_Chanial_2008} and \citet{Galametz2009} galaxies are finally obtained from the \iras\ Faint Source catalogue and the \iras\ catalog of Point Sources (IPAC 1986). The choice of the \iras\ data is of little influence on the dust masses we derive from the SED models. {\revisedbis  We finally note that there is no systematic difference between the wavelength coverage of dwarf galaxies compared to higher metallicity galaxies. }

\subsection {The building blocks of our fiducial SED model}

The SED of a galaxy synthesizes the emission at each wavelength of its different components: H~{\sc ii} regions, molecular clouds, diffuse ISM etc. The SED fits obtained from any model are dependant on the assumptions we make {\it a priori} about parameters such as radiation field or dust properties. Despite this dependance, valuable comparisons can be made from the study of different galaxies using the same SED modelling method. 

{\revised We first consider that the sources of IR emission are the dust grains (PAHs, silicate, graphite) and the stellar continuum. Our SED modelling approach uses realistic intrinsic dust properties and is commonly used in the litterature \citep{Dale_Helou_2002, Paradis2009, Whaley2009}. The properties of silicate and grapite grains, namely their size distribution and chemical composition, are the standard and up-to-date properties of the Bare-Gr-S dust model \citet{Zubko2004}. We emphasis that this model is consistent with the recent results of \citet{Draine2007} and the DustEM model of \citet{Compiegne2011}. }
We assume that the dust particles are thus a mixture of grains with solar depletion constraints. The optical properties of the silicate and graphite grains are taken from \citet{Weingartner2001} and \citet{Laor1993} respectively \footnote{We also refer to \citet{Draine_Li_2007} and http://www.astro.princeton.edu/$\sim$draine/dust/dust.html where dust properties can be found.}. In order to better constrain the SED in the mid-IR range, we introduce two free parameters to model the PAH emission: the PAH charge fraction (f$_{PAH+}$) and the PAH-to-dust mass ratio (f$_{PAH}$), normalised to the Galactic value of 0.046. The PAH and PAH+ properties (e.g. the absorption efficiencies) are taken from \citet{Draine_Li_2007}. {\revisedbis When observations are lacking 8 \mic\ measures and to avoid a degeneracy in the modelling, this f$_{PAH}$ parameter is fixed to 0.5 (which means half the Galactic value).} This choice does not significantly influence the total dust mass derived from the models. The total mass of dust M$_{dust}$ is a free parameter of our modelling. \\ 

\begin{figure*}
\centering
\begin{tabular}{m{7.5cm}m{7.5cm}m{7.5cm}}
	\includegraphics[width=7.5cm ,height=5.5cm]{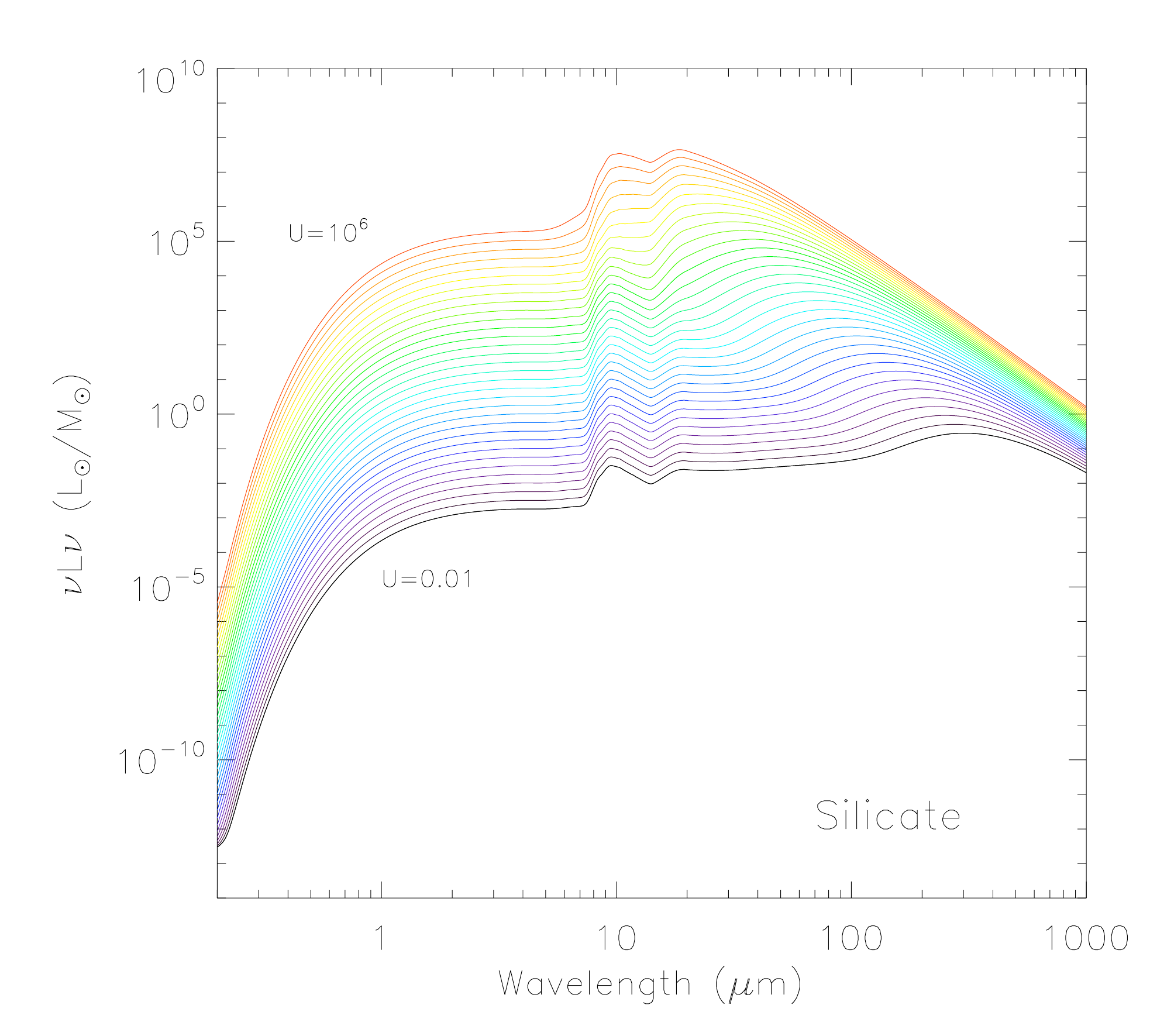} &
	\includegraphics[width=7.5cm ,height=5.5cm]{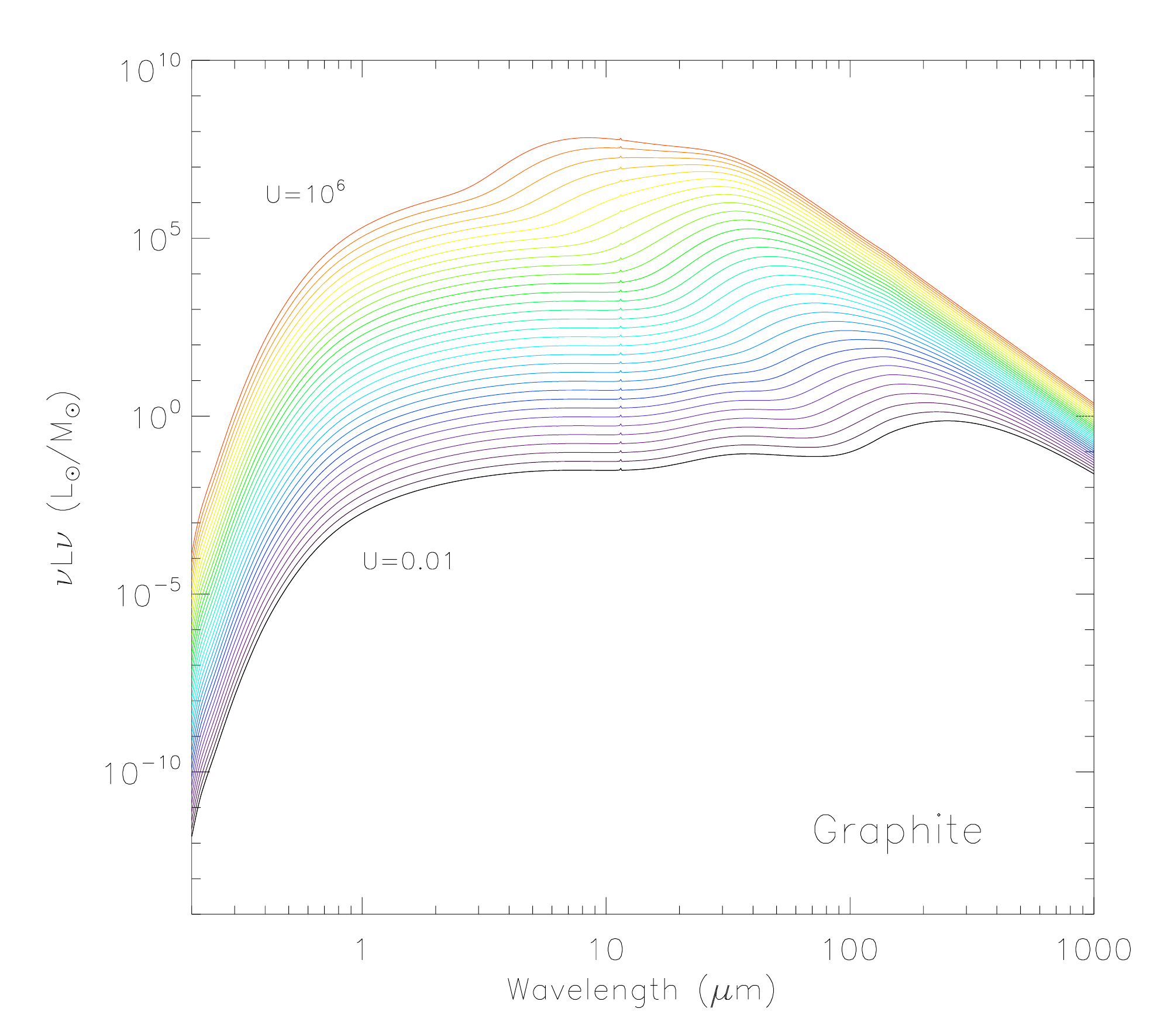} \\
	\includegraphics[width=7.5cm ,height=5.5cm]{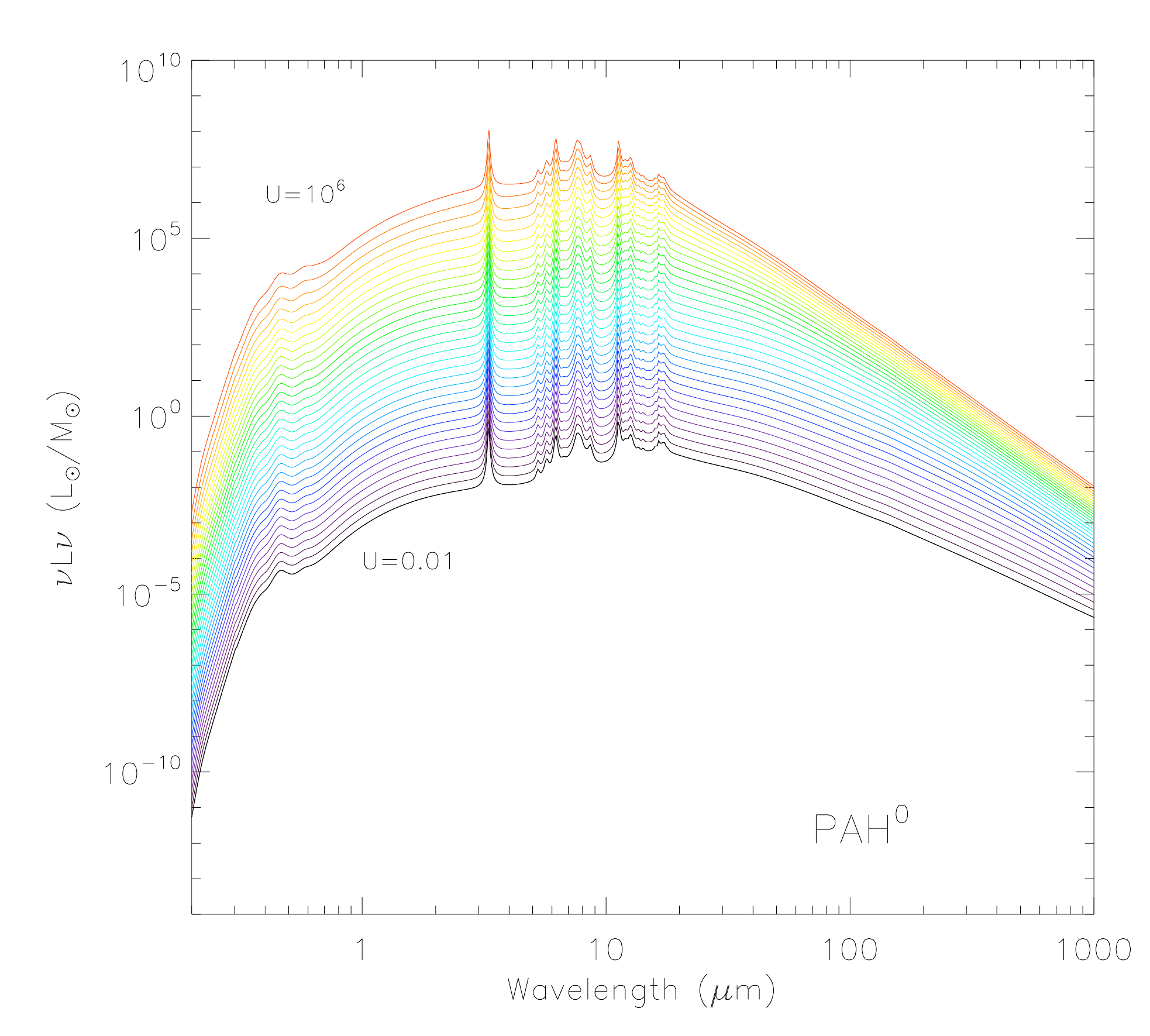} &
	\includegraphics[width=7cm ,height=5cm]{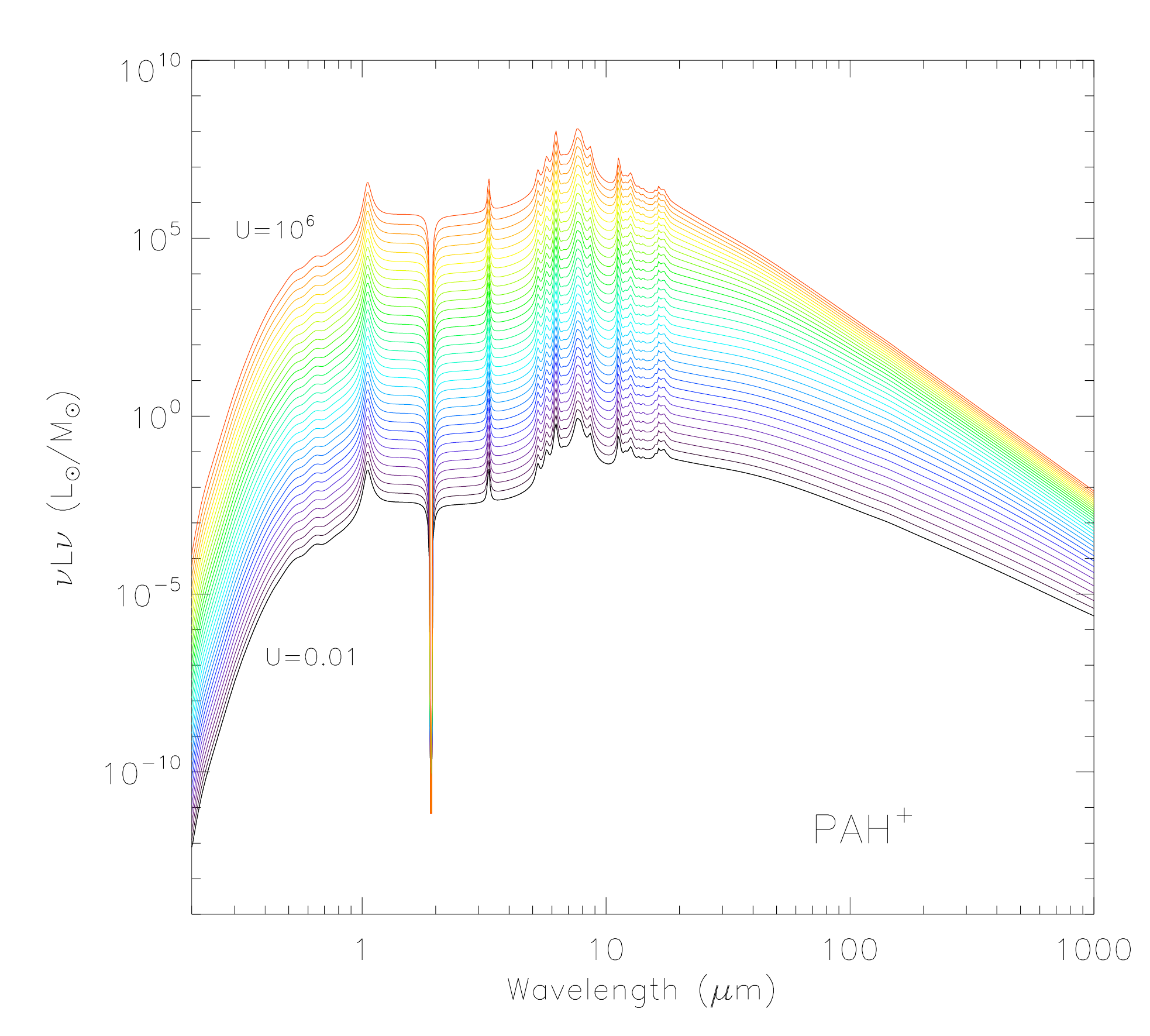} 
\end{tabular}
\caption[Emission of dust grains for different U]{Emission of silicate, graphite, neutral and ionized PAHs for different values of {\it U}.}
\label{nulnu_U}
\end{figure*}

We adopt the spectral shape of the Galactic diffuse ISM Interstellar Radiation Field (ISRF) determined by \citet{Mathis1983}. Previous studies have shown that the form of the global ISRF is different from normal or dusty galaxies to low-metallicity galaxies \citep[e.g.][]{Madden2006}, an effect attributed to the lower dust attenuation and thus the larger mean free path length of the ionizing photons in low-metallicity ISM and the predominance of massive stars in those environments. Nevertheless, the shape of the ISRF does not influence the far-IR to submm regime of the SED and thus the total dust mass derived \citep[see][]{Galametz2009}. The radiation field indeed mainly affects the emission of the non-equilibrium grains that do not significantly contribute to the dust mass.  

The SED is multiplied by a dimensionless scaling factor {\it U}, with {\it U}=1 corresponding to a radiation field of the local solar neighborhood value of 2.17 $\times$ 10$^{-5}$ W m$^{-2}$ \citep{Mathis1983}. 
{\revised To describe the various environments within which the various components of the ISM reside, we combine the individual SEDs using the prescription defined in \citet{Dale2001}. This parametrisation of the analytical form of the temperature distribution has been widely adopted to describe the ISRF in galaxies, for example by \citet{Dale_Helou_2002}, \citet{Galliano_Dwek_Chanial_2008}, \citet{Paradis2009}, among others.
It assumes a power-law distribution of the dust mass over a wide range of interstellar radiation heating intensities and conservatively links the hot and cold emission in galaxies: \\}

\begin{equation}
dM_{dust}({\it U}) \propto {\it U} ^{-\alpha}dU 	
\end{equation}

\noindent where M$_{dust}$ is the dust mass heated by the radiation field, U its intensity and $\alpha$ the coefficient of the power law describing the contributions of the individual SEDs (also a free parameter of our model). \\

Radiation field intensities are ranging between the two free parameters U$_{min}$ and U$_{max}$. The behavior of the emission of the different dust components as a function of U are plotted in Fig.~\ref{nulnu_U}. {\revised We note that \citet{Draine2007} used a similar parametrisation but add a "diffuse ISM" component with a unique intensity factor U=U$_{min}$. They choose to restrain this parameter to values $\ge$ 0.7 as an {\it ad hoc} measure to avoid invoking large dust masses when the submm slope of the SED is not constrained. This assumption is further discussed in $\S$ 4.1.\\}

Finally, we subtract the stellar contribution to the SED in the mid-IR using 2MASS and the \spitz/IRAC 3.6 \mic\ bands. We model the contribution of old stars using the stellar evolution program PEGASE \citep{Fioc_Rocca_1997}, assuming that the stellar population has undergone an instantaneous burst 5 Gyr ago. We adopt the Salpeter Initial Mass Function : N(M) dM $\propto$  M$^{-1.35}$ dM \citep{Salpeter1955}, and an initial metallicity of Z=Z$_\odot$. The mass of old stars M$_{oldstars}$ is a free parameter of our modelling. The choice of the IMF does not particularly influence our study since we are only fitting the IR and do not want to necessarily quantify the mass of old stars. We finally note that no radio component was introduced in the modelling compared to the similar model of \citet{Galliano_Dwek_Chanial_2008}.\\

In summary, the free parameters of our modelling are: 

\begin{itemize}
\item {\bf M$_{dust}$} the total mass of dust 
\item {\bf f$_{PAH}$} the PAHs-to-dust mass ratio (normalised to the Galactic value) 
\item {\bf f$_{PAH+}$} the ionized PAHs-to-total PAHs mass ratio
\item {\bf $\alpha$} the index describing the fraction of dust exposed to a given intensity
\item {\bf U$_{min}$} the minimum  heating intensity
\item {\bf U$_{max}$} the maximum heating intensity
\item {\bf M$_{oldstar}$} the mass of old stars 
\end{itemize}

The overall modelling is an iterative process. We compute the temperature distribution of the dust grains heated by the absorption of the stellar radiation. The IR emission is then precalculated for each dust grain (different sizes, different properties) and each intensity U. The sum of these discrete contributions leads to a global SED model of the galaxy:

{\small \begin{eqnarray}
L^{tot}_{\nu} (\lambda) & = & L^{PAH+dust}_{\nu} (\lambda) + L^{star}_{\nu}(\lambda) \nonumber \\
			   & = & \frac{1-\alpha}{U_{max}^{1-\alpha}-U_{min}^{1-\alpha}}\int_{U_{min}}^{U_{max}} l^{PAH+dust}_{\nu}(U,\lambda)U^{-\alpha}M_{dust}dU+L^{star}_{\nu}(\lambda) 
 \end{eqnarray}}
 
 with {\small \begin{eqnarray}
l^{PAH+dust}_{\nu}(U,\lambda)  & = &  f_{PAH}[L^{PAH^0}_{\nu}(U,\lambda)(1-f_{PAH+})+ L^{PAH+}_{\nu}(U,\lambda)f_{PAH+}] \nonumber \\
					  & + &  L^{graphites}_{\nu}(U,\lambda)+L^{silicates}_{\nu}(U,\lambda)	   
 \end{eqnarray}}
 
Both \iras\ and \spitz\ bands are color-corrected. The \iras\ fluxes are corrected using a $\nu$ $\times$ F$_{\nu}$ convention. The \spitz\ bands are corrected using the convention described on the \spitz\ website \footnote{IRAC: http://ssc.spitzer.caltech.edu/irac/iracinstrumenthandbook/21/ MIPS: http://ssc.spitzer.caltech.edu/mips/mipsinstrumenthandbook/51/}.  \\

Finally, all data are weighted with respect to the density of their neighbours. The model obtained is integrated within each filter and compared to observed values using a $\chi$$^{2}$ minimisation based on the Levenberg-Marquardt algorithm (least squares fitting routines):

\begin{equation}
{\chi^2}=\sum_i\left(\frac{ \lambda_{i+1}-\lambda_{i-1}}{2\lambda_i}\right)\left[\frac{ L_{\nu}^{obs}(\lambda_i)-L_{\nu}(\lambda_i)}{\Delta L_{\nu}^{obs}(\lambda_i)/2}\right]^2
 \end{equation}
\noindent where $\Delta$ L$_{\nu}^{obs}$($\lambda$$_i$) is the error on the luminosity at a given wavelength. \\


\begin{figure}
    \centering
    \begin{tabular}{m{8.1cm}}
      \includegraphics[width=8.5cm ,height=5.5cm]{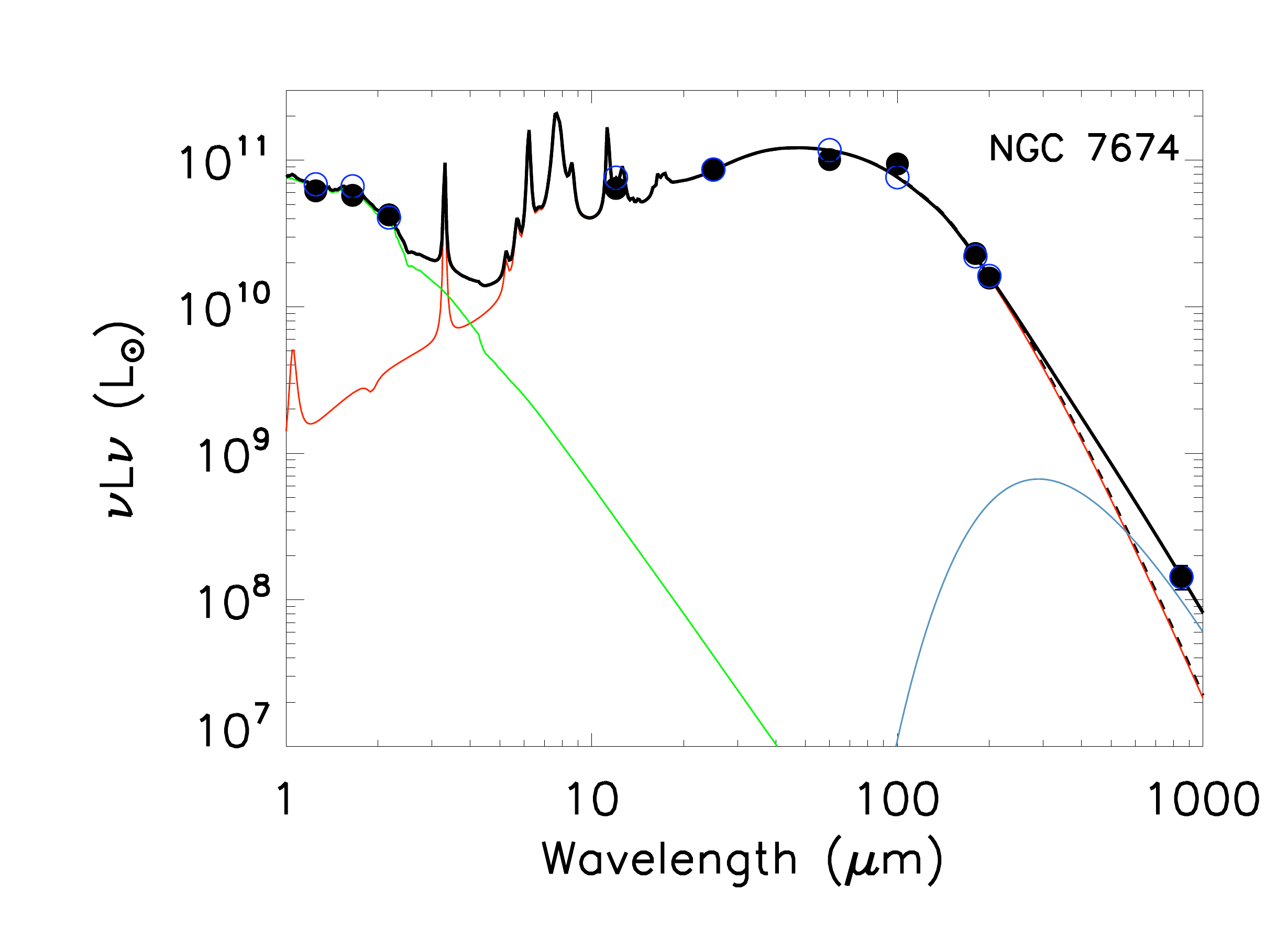}    \\
       \includegraphics[width=8.5cm ,height=5.5cm]{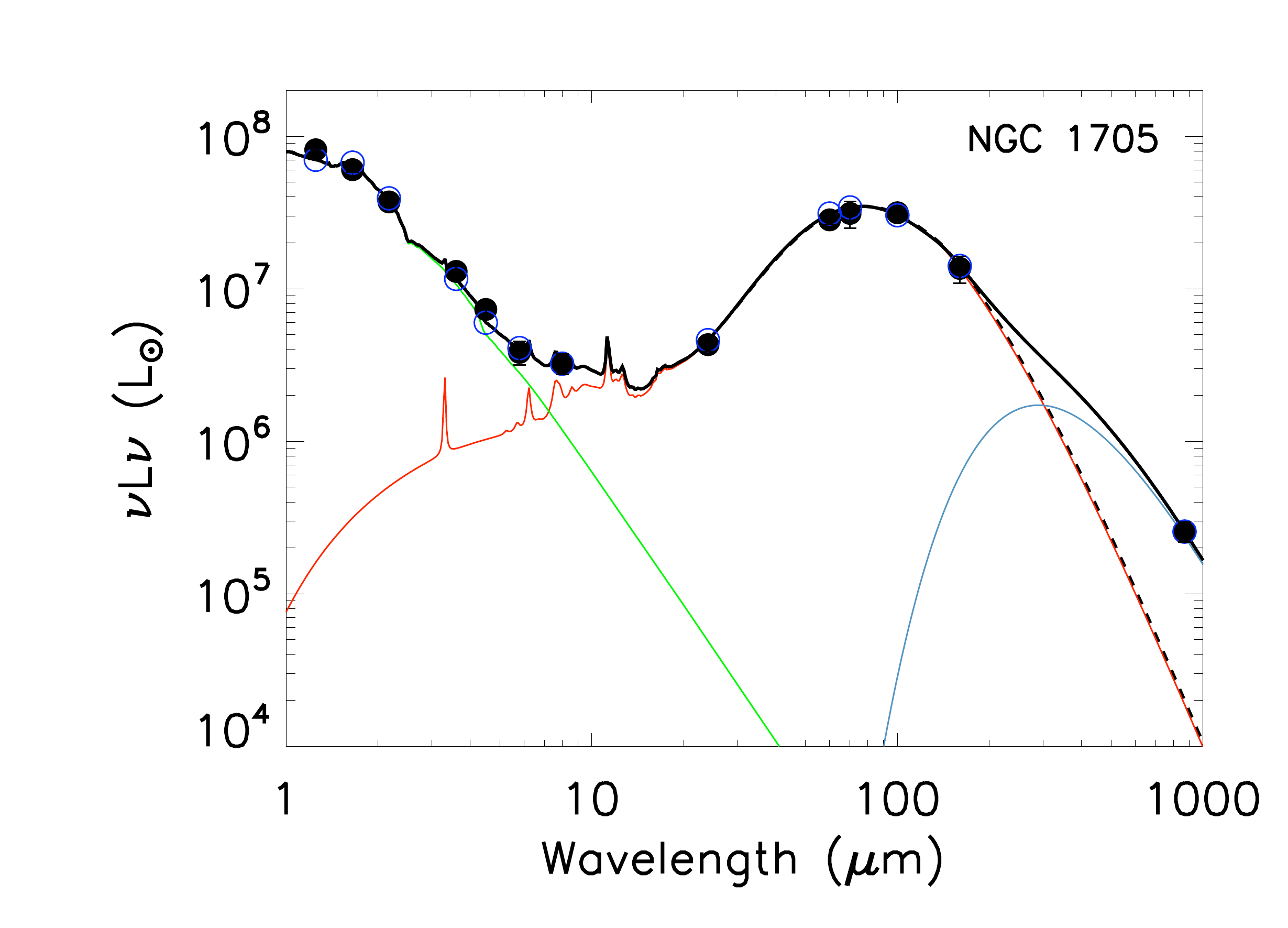}   \\
      \includegraphics[width=8.5cm ,height=5.5cm]{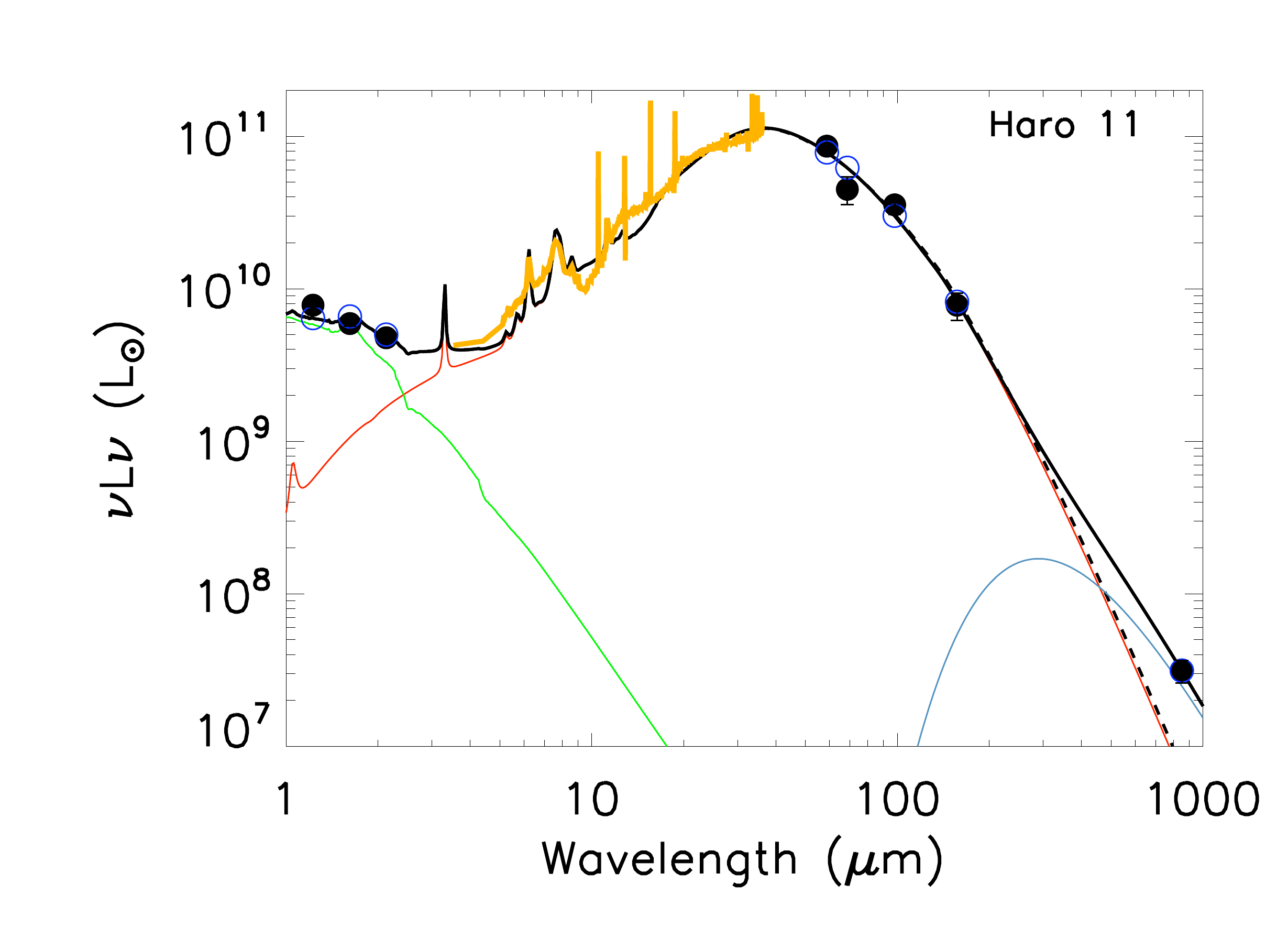} \\
       \end{tabular}
    \caption{SED models of the galaxies for which an excess at submm wavelengths is detected. The SEDs modelled with submm data are indicated by the black lines while the dashed black lines indicates the model obtained when submm data are omitted in the modelling. Observational constraints are superimposed (filled circles). When the error bars are not shown, the errors are smaller than symbols. The thick orange line shows the IRS spectrum used in the SED modelling when available. The open circles indicate the expected modelled fluxes integrated over the instrumental bands. The green, red and blue lines respectively distinguish the stellar, the warm dust and the cold dust contributions to the SED determined with submm data.}
    \label{SED_CD}
\end{figure}


\addtocounter {figure}{-1}
\begin{figure*}
    \centering
    \begin{tabular}{m{8.1cm}m{9cm}}
      \includegraphics[width=8.5cm ,height=5.5cm]{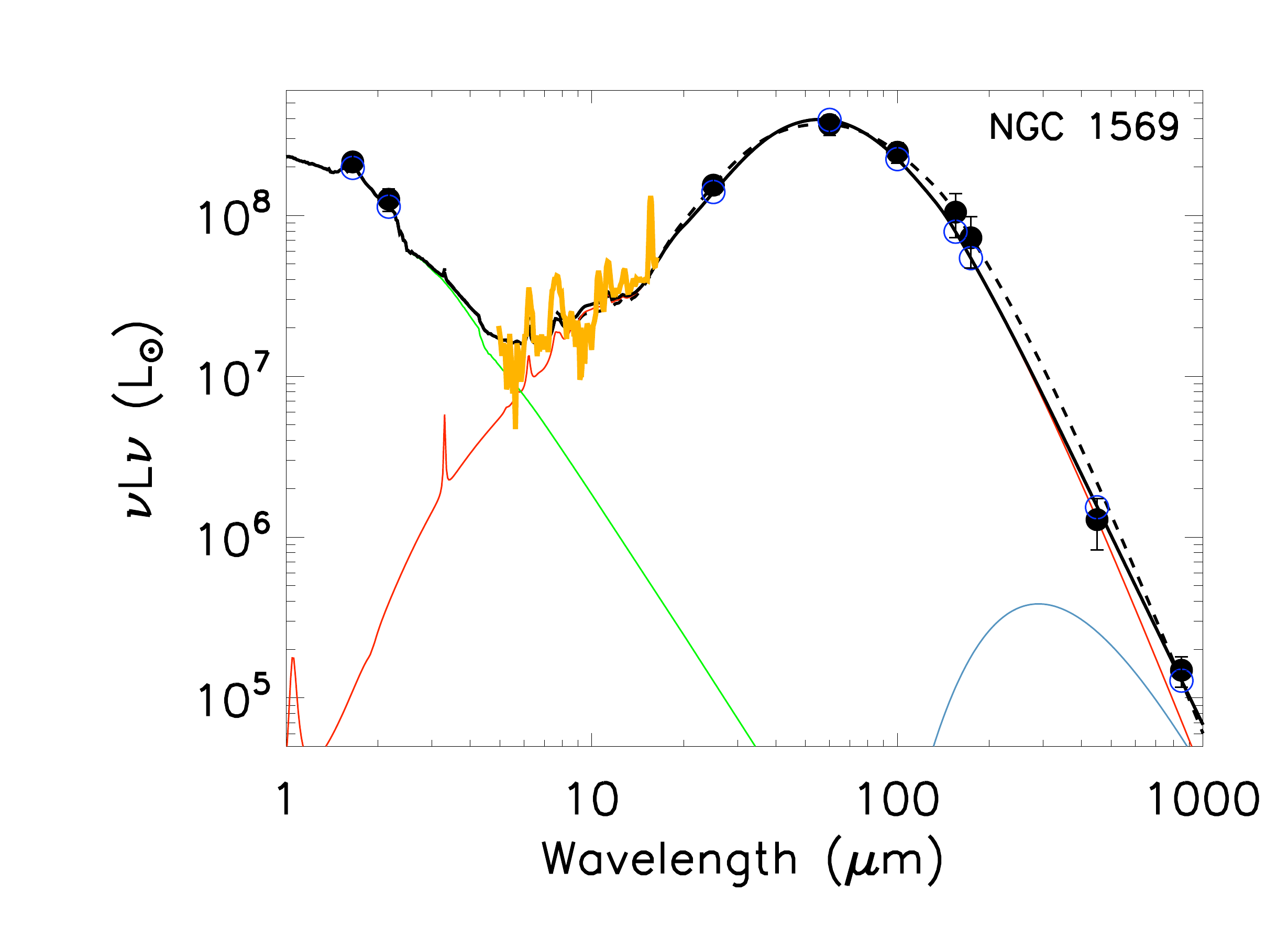}  &  \includegraphics[width=8.5cm ,height=5.5cm]{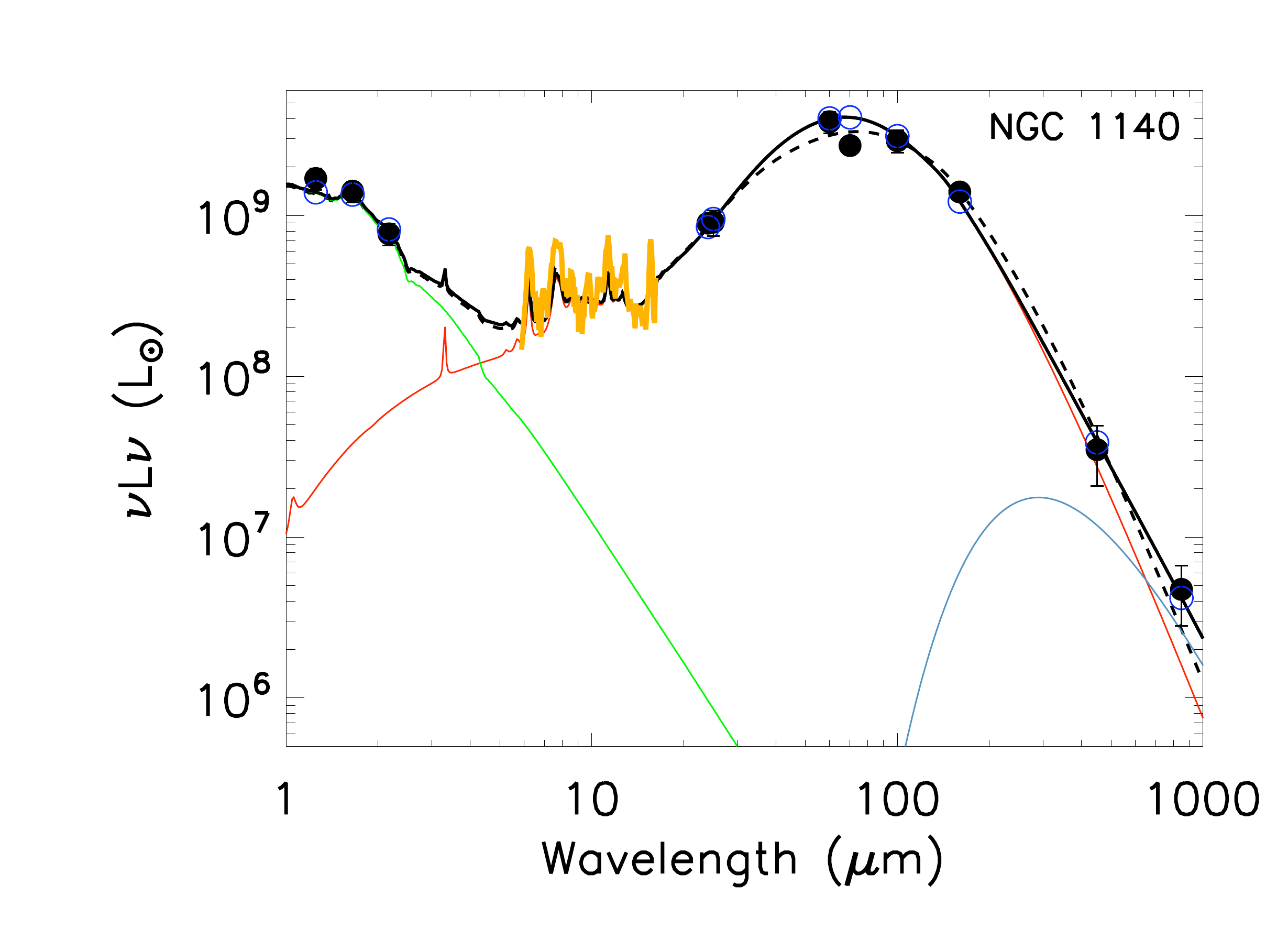} \\
    \includegraphics[width=8.5cm ,height=5.5cm]{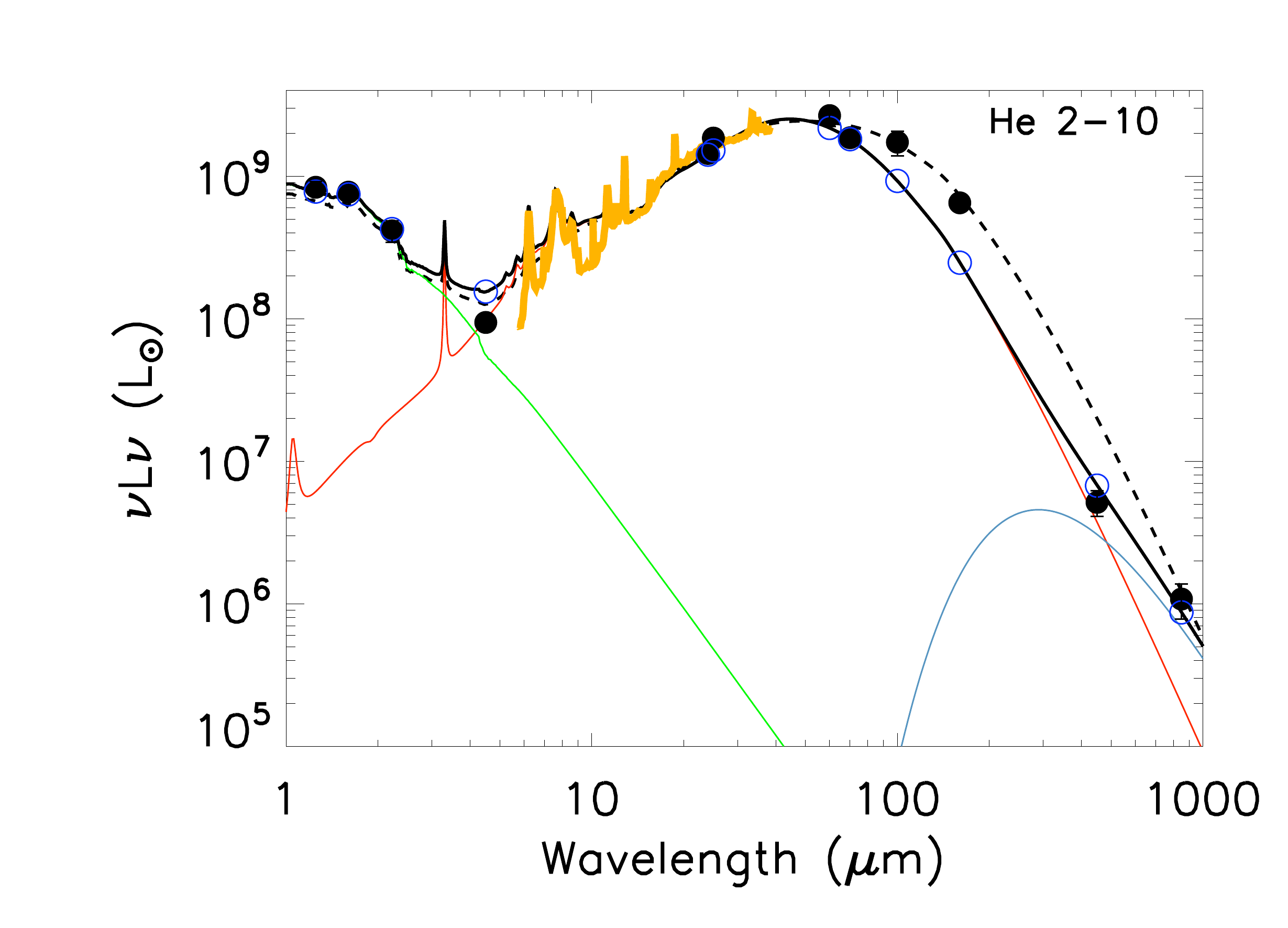} & \includegraphics[width=8.5cm ,height=5.5cm]{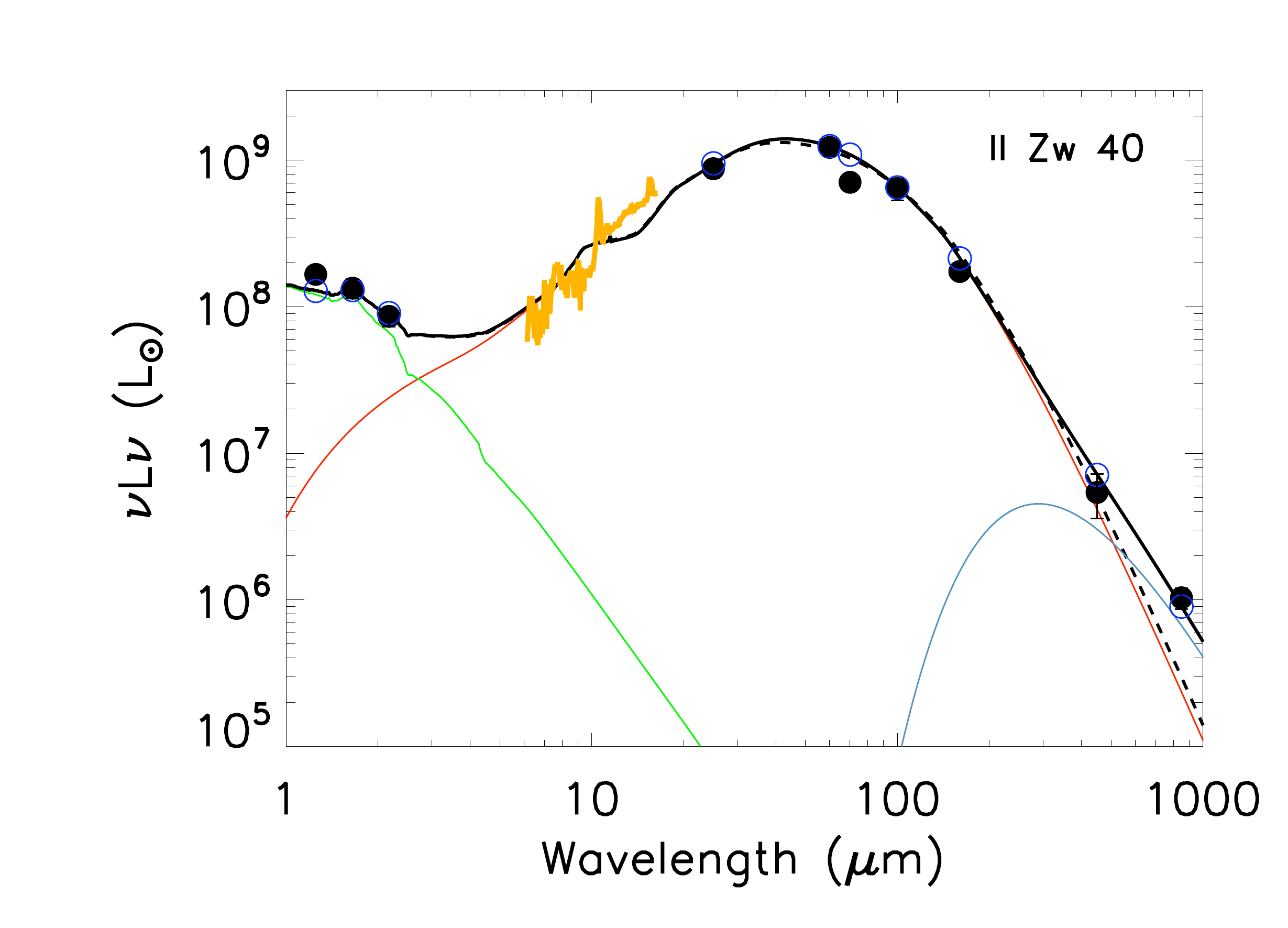}\\
       \includegraphics[width=8.5cm ,height=5.5cm]{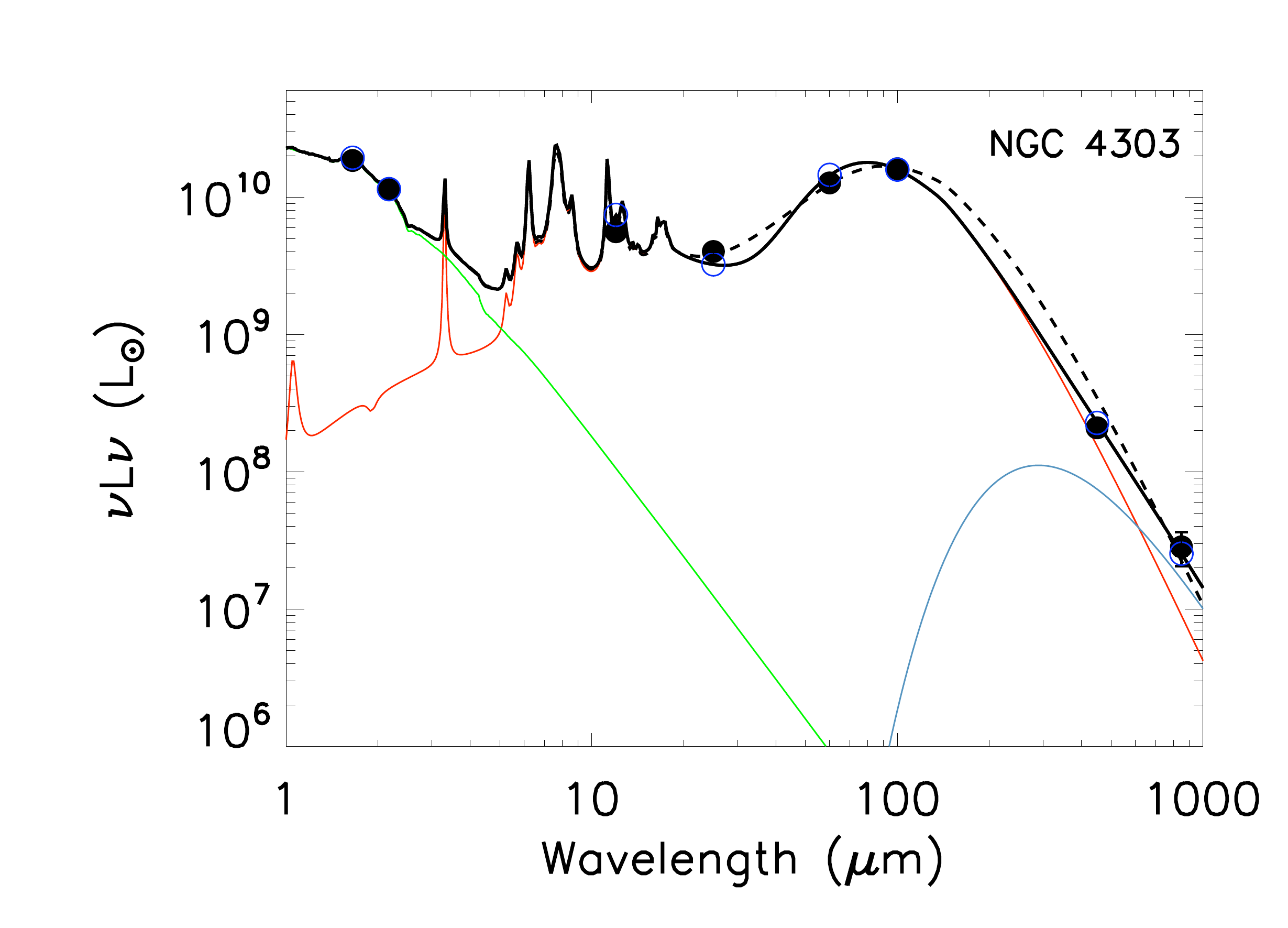} & \includegraphics[width=8.5cm ,height=5.5cm]{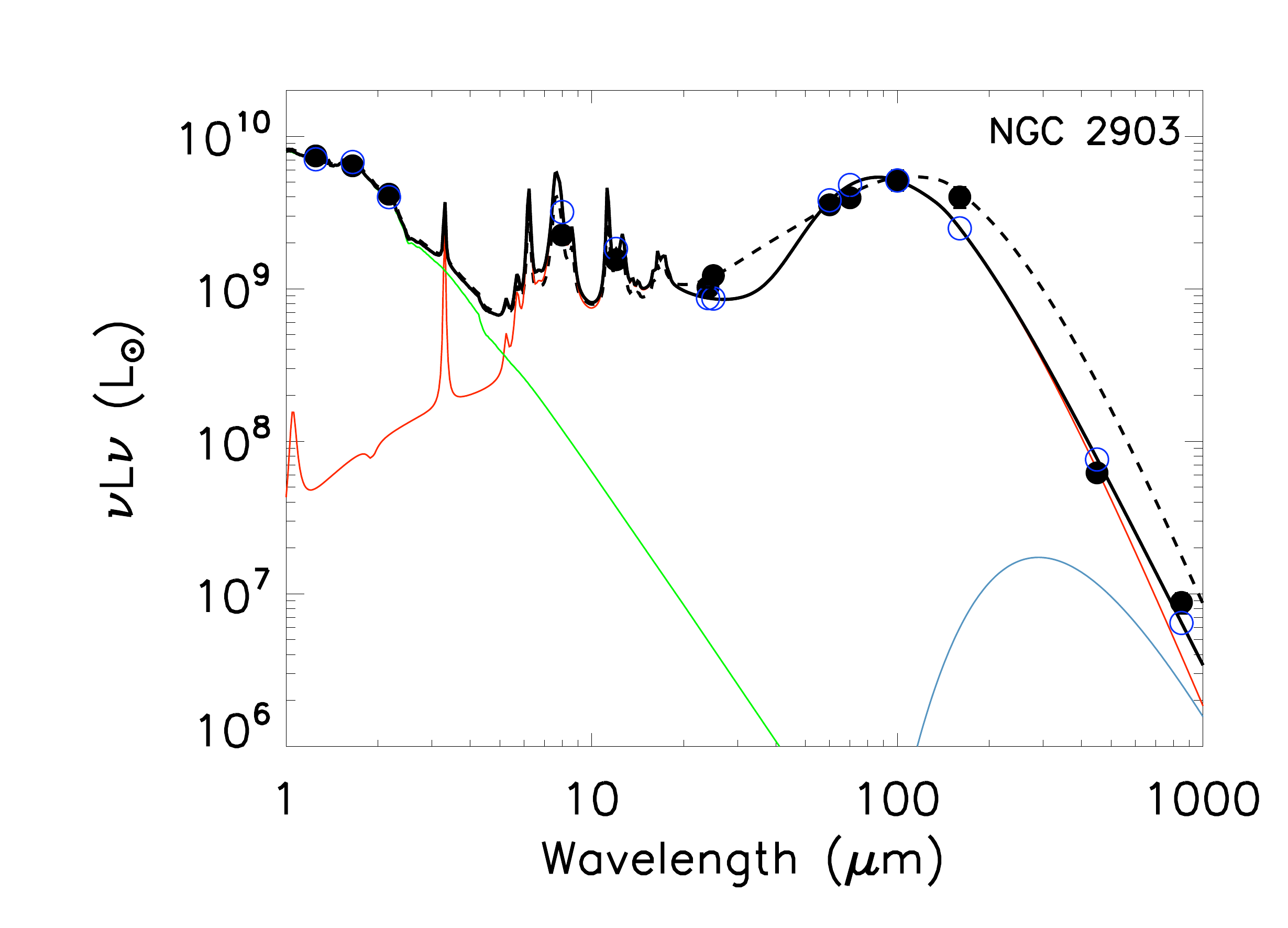}  \\
       \end{tabular}
    \caption{ continued }
    \label{SED_CD}
\end{figure*}

\subsection {Modelling the submm excess}

{\revisedbis Most of the galaxies of our sample can be fit with our SED model up to submm wavelengths. Nevertheless, for 9 galaxies (namely NGC~1569, NGC~1140, He~2-10, II~Zw~40,  NGC~4303, NGC~2903,  NGC~7674, NGC~1705 and Haro~11), our model can not fit together the far-IR measures and the elevated submm fluxes and an excess at submm wavelengths is detected. 
Those galaxies are mostly low-metallicity dwarf galaxies. Such excess is often reported while modelling the SEDs of low-metallicity galaxies \citep{Galliano2003,Dumke2004, Galliano2005,Bendo2006,Marleau2006, Galametz2009, Grossi2010, OHalloran2010, Israel2010,Bot2010}.

{\revisedbis We note that \citet{Li_Draine_2001} modified the standard silicate emissivity at $\lambda$ $>$ 250 \mic\ in order to match the average high Galactic latitude dust emission and the Galactic ``submm excess" observed with DIRBE/FIRAS. In this paper, we use the \citet{Zubko2004} model that fits the same constraints. The submm excess detected in our 9 galaxies may have the same origin as the Galactic submm excess, but the amplitude of the excess observed in the Milky Way is much smaller than observed in dwarf galaxies.  }


This paper will assume that the excess is due to  cold dust and explore the consequences of this hypothesis on the global properties (total dust mass and D/G) of those 9 galaxies. }
For these 9 galaxies, we thus add an independent thermal cold dust component to our fiducial model. The luminosity of the cold dust component is calculated using: 

\begin{eqnarray}
L_{\nu}^{CD}(\lambda)  & = & M_{CD} ~ \kappa_{\nu,CD} ~ \left(\frac{\lambda_o}{\lambda}\right)^{\beta} ~ (4 \pi B_{\nu}(\lambda,T_{CD})) \nonumber \\
		& = & M_{CD} ~ \frac{( \pi a^2 Q_o )}{(4/3~\pi a^3 \rho)} ~ \left(\frac{\lambda_o}{\lambda}\right)^{\beta} (4 \pi B_{\nu}(\lambda,T_{CD})) \nonumber \\
		& = & M_{CD} ~ 3 \pi ~ \left(\frac{1}{\rho}\right) ~\left(\frac{Q_o}{a}\right) ~ \lambda_o^{\beta} ~ \left[\lambda^{- \beta} B_{\nu}(\lambda,T_{CD})\right]
\end{eqnarray}


\noindent where $B_{\nu}$ is the Planck function, T$_{CD}$ the cold dust temperature, $\beta$ the emissivity coefficient. {\revisedbis We adopt $\lambda$$_0$=100 \mic\ for the reference wavelength. We assume graphitic values for the mass density of the grains $\rho$=2.5 $\times$ 10$^3$~kg~m$^{-3}$   \citep[close to the mass density of carbonaceous grains of 2.2 $\times$ 10$^3$~kg~m$^{-3}$ of][] {Draine_Li_2007} and the absorption coefficient Qo/a =150 cm$^{-1}$ at $\lambda$$_0$}, with a, the radius of the grain. We are studying an effect which is difficult to constrain completely with one or two submm fluxes. 

The lack of constraints forces us to fix the temperature and the dust emissivity index of this blackbody to conservative values.  \\

{\it Temperature - } ~\citet{Galliano2003, Galliano2005} and \citet{Marleau2006} suggest that their submm excess observed in low-metallicity galaxies could be explained by the presence of cold dust distributed in small and dense clumps and modelled the excess with an independant cold dust component with  temperatures between 5 and 10 K. {\revisedbis We tried modeled the 9 galaxies with a cold dust component with T$>$10K (ex:15K). In most cases, this leads to fits with higher chi-square values. When the 450 \mic\ measure is available, the model has difficulty to fit the peak of the SED and the 450 \mic\ data together and usually leads to a SED model that overestimates the 450 \mic\ data.} The temperature choice influences the dust mass derived from our modelling and we decide to fix the cold dust component temperature to 10K. Using 10K (and not $<$10K) dust also prevents us from requiring excessive cold dust. \\

{\revisedbis {\it Emissivity - } ~A flattening of the submm slope is often observed when 450 + 850 or 870 \mic\ measurements are available (see Fig.~\ref{SED_CD}). As suggested by several hypothesis to explain the submm excess and reviewed in the introduction of this paper, an increase of the emissivity at submm wavelengths is expected to explain the flattening of the submm slope. Moreover, \citet{Galliano2003,Galliano2005} and \citet{Galametz2009} modeled the submm excess they detected with a very cold dust component and found that an emissivity index of 2, as commonly used to describe the submm slope of SEDs \citep[i.e.][]{Dunne_Eales_2001,Draine_Li_2007} leads to unrealistically high dust masses. \citep[For instance, ][ found that dust masses were multiplied by a factor of 6-7 when using $\beta$=2 in lieu of 1]{Galametz2009}. We conservatively adopt the value $\beta$=1.} We fix our absorption to the model of \citet{Li_Draine_2001}, which may deviate when a dust emissivity index of 1 (and not the standard emissivity of 2) is used. We assume that this value would still be applicable here. \\

\begin{landscape}
\begin{table}
\caption[Description of our broad sample and dust mass estimates]{Description of the sample and dust mass estimates with and without submm data}
\label{Effects_of_the_submm_Table}
 \centering
 \begin{tabular}{p{1.8cm}p{1.2cm}p{0.8cm}p{0.5cm}p{0.8cm}p{0.5cm}p{1cm}p{0.5cm}p{0.8cm}p{0.8cm}p{1.2cm}p{1.2cm}p{0.8cm}p{1cm}p{1.2cm}p{1cm}p{1.2cm}}
\hline
\hline
(1) &  (2) &\multicolumn{2}{c} {(3)} & \multicolumn{2}{c} {(4)} &\multicolumn{2}{c} {(5)} & \multicolumn{2}{c} {(6)} &\multicolumn{3}{c} {(7)} &\multicolumn{2}{c} {(8)} & \multicolumn{2}{c} {(9)}\\
&&&&&&&&&&&&&&&\\
Name& Distance &\multicolumn{2}{c} {12+log(O/H)} &\multicolumn{2}{c} {S$_{450}$} &\multicolumn{2}{c} {S$_{850}$} & \multicolumn{2}{c} {log HI}&\multicolumn{3}{c} {log H$_2$} & \multicolumn{2}{c} {log M$_{dust~}$$_{w/o~submm}$} & \multicolumn{2}{c} {log M$_{dust~}$$_{with~ submm}$}\\
&&&&&&&&&&&&&&&&\\
 &           & Value & Ref. & Value & Ref. & Value & Ref. & Value &Ref. & {\revisedbis Value $_{Gal}$} & {\revisedbis Value $_{B02}$} & Ref. & Value   & unc                           & Value       & unc    \\
 &(Mpc) &            &          & 	(Jy)    &         &   (mJy) &        &(\msun)&        &  (\msun)    &(\msun)                &          & (\msun)& $\%$~M$_{dust}$ &  (\msun)   &    $\%$~M$_{dust}$ \\
\hline
&&&&&&&&&&&&&&&&\\
NGC 337			&	24.7	&	-	&	-	&	-	&	-	&	350		&	11	&	- 	&	-	& -        &	- 	&	-	&	8.29		&	86$\%$ 		&	7.7		& 23$\%$	\\
NGC 2798		&	24.7	&	-	&	-	&	-	&	-	&	190		&	11	& 9.29	&	13	& 9.53 &	- 	&	14	&	7.22 		&	56$\%$		&	7.23 		& 26$\%$ \\
NGC 4631		&	9	&	-	&	-	&	30.7	&	11	&	5730		&	11	& 10.09	&	13	& 9.25 &	- 	&	14	&	9.36 		& 	60$\%$		&	7.96 		& 22$\%$ \\
NGC 5195		&	8.2	&	-	&	-	&	-	&	-	&	260		&	11	& -		&	-	& 8.31 &	- 	&	14	&	6.43 		&	97$\%$		&	6.84 		& 39$\%$\\
NGC 5713		&	26.6	&	-	&	-	&	-	&	-	&	570		&	11	& 9.93	&	13	& 9.78 &	- 	&	14	&	8.31 		& 	44$\%$		&	7.91  	& 31$\%$ \\
NGC 5866		&	12.5	&	-	&	-	&	0.8	&	11	&	140		&	11	& $<$8.28&	13	& 8.69 &	- 	&	14	&	6.79 		&	24$\%$		&	6.50  	& 10$\%$\\
NGC 520			&	27	&	- 	&	-	&	-	&	- 	&	325		&	2	& 9.54	&	20	& 9.42 &	- 	&	21	&	7.55 		&	49$\%$		&	7.56 		& 66$\%$	\\
NGC 6240		&	98	&	- 	&	-	&	1.0	&	4 	&	150		&	4	& 9.95	&	22	& 9.57 &	- 	&	22	&	8.25		&	19$\%$		&	8.36		& 28$\%$\\
NGC 4569		&	20	&	9.3 	& 	64	&	-	&	- 	&	470		&	11	& 8.80	&	13	& 9.73 & 9.25   &	14	&	8.84		&	25$\%$		&	7.56  	&17$\%$	\\
M83	 			&	4.5	&	9.2 	&	47	&	-	&	- 	&	-		&	-	& 9.71	&	15	& 9.59 & 9.21	&	16	&	8.31 		&	30$\%$		&	6.93 		&48$\%$	\\
NGC 1808		&	11	&	9.1   &	48	&	8.13	&	9 	&	1300		&	9	& 9.25	&	18	& 9.30 & 9.03	&	19	&	7.53 		&	53$\%$		&	7.49 		& 61$\%$	\\
NGC 7552		&	22.3	&	9.0	&	64	&	-	&	- 	&	800		&	11	& 9.68	&	14	&     -    &    -		&	-	&	8.11 		&	94$\%$		&	7.87 		&21$\%$	\\
NGC 1097		&	12	&	9.0   &	49	&	-	&	- 	&	1440		&	11	& 9.71	&	23	& 8.67 & 8.50	&	24	&	8.07 		&	47$\%$		&	7.71		&17$\%$\\
M82				&	3.6	&	9.0   &	50	&	49	&	1 	&	-		&	-	& 8.95	&	25	& 9.20 & 9.03	&	26	&	6.66 		&	19$\%$		&	6.89 		&16$\%$	\\
NGC 1068		&	15	&	9.0 	&	51	&	-	&	- 	&	-		&	-	& 9.35	&	27	& 9.92 & 9.75	&	28	&	9.52		&	88$\%$		&	8.17 		& 42$\%$	\\
NGC 2903 		&	6.3	&	8.94 &	46	&	7.94	&	9 	&	2120		&	5	& 8.98	&	5	& 9.06& 8.95	&	28	&	 7.62 		&	 34$\%$		&	 6.86  $^b$&  41$\%$ 	\\
NGC 4536		&	25	&	8.9 	&	64	&	-	&	- 	&	420		&	11	& 9.71	&	13	& 9.77 & 9.70	&	13	&	9.41 		&	98$\%$		&	7.71 		&18$\%$	\\
NGC 891			&	9.6	&	8.9   &	45	&	-	&	- 	&	-		&	-	& 9.88	&	22	& 9.82 & 9.75	&	22	&	7.69 		&	23$\%$		&	7.66  	& 25$\%$	\\
He 2-10 			&	8.7	&	8.9   &	52	&	0.34	&	8	&	130		&	8	& 8.49	&	29	& 8.26 & 8.19	&	29	&	 6.28		&	 37$\%$		&	 5.74 	$^b$	& 46$\%$	\\
IC 342			&	3.8	&	8.9  	&	46	&	-	&	- 	&	-		&	-	& 10.1	&	22	& 9.83 & 9.76	&	22	&	8.09 		&	21$\%$		&	$<$ 7.27 	& -	\\
MCG+02-04-025	&	122	&	8.87 &	5	&	-	&	- 	&	390  		&	5	& 9.24	&	5	& -        &  -	&	-	&	7.69 		&	33$\%$		&	7.85 		& 60$\%$	\\
NGC 4303 		&	15.2	&	8.84  &	46	&	4.6	&	9 	&	1180		&	5	& 9.42	&	5	& 9.70 & 9.70	&	28	&	7.87 		&	15$\%$		&	7.32 	$^b$	&4$\%$ 	\\
NGC 7469		&	64	&	8.8   &	53	&	-	&	- 	&	2640		&	5	& 9.18	&	5	& 9.96 & 9.99	&	5	&	8.84 		&	31$\%$		&	8.07 		&3$\%$	\\
NGC 5256		&	109	&	8.75 &	5	&	-	&	- 	&	820  		&	5	& 0.00	&	5	&10.28& 10.36	&	5	&	8.21		&	10$\%$  		&	8.22 		&23$\%$	\\
NGC 5953		&	26	&	8.73 &	5	&	-	&	- 	&	1820		&	5	& 8.76	&	5	& 9.27 & 9.37	&	5	&	7.34		&	19$\%$  		&	7.28 		&14$\%$	\\
NGC 6946		&	5.5	&	8.7   &	46	&	18.5	&	11 	&	1200		&	11	& 9.49	&	5	& 9.40 & 9.53	&	28	&	7.30		&	6$\%$		&	7.05 		&1$\%$	\\
M51				&	8.4	&	8.7 	&	54	&	-	&	- 	&	15000	&	10	& 9.70	&	30	& 9.81& 9.94	&	28	&	8.73		&	97$\%$		&	8.51 		&34$\%$	\\	
NGC 3995		&	43	&	8.66 &	5	&	-	&	- 	&	1260		&	5	& 9.79	&	5	& -       & 	-	&	-	&	7.57		&	57$\%$		&	7.73 		&28$\%$	\\
NGC 3994		&	41	&	8.61 &	5	&	-	&	- 	&	1060		&	5	& 9.45	&	5	& -        &    -	&	-	&	7.48		&	36$\%$		&	7.50 		&37$\%$	\\
NGC 6052		&	62	&	8.6   &	5	&	0.72 &	3 	&	950  		&	5	& 9.58	&	5	& 9.49 & 9.72	&	5	&	7.73		&	15$\%$		&	7.76 		&24$\%$	\\
NGC 4826		&	5.6	&	8.59 	&	44	&	-	&	- 	&	1230		&	11	&  8.49	&	14	& 8.79 & 9.03 	&	14	&	7.73		&	84$\%$		&	6.84 		&44$\%$	\\
NGC 1222		&	32	&	8.57 &	55	&	-	&	- 	&	840  		&	5	& 9.08	&	5	& -        &      -	&	-	&	6.93		&	13$\%$  		&	7.10 		&22$\%$	\\
NGC 7674		&	113	&	8.56 &	5	&	-	&	- 	&	1080		&	5	& 10.03	&	5	&10.29& 10.56	&	5	&	7.82		&	6$\%$		&	8.07 	$^b$	&13$\%$	\\
NGC 7714		&	37	&	8.5 	&	56	&	-	&	- 	&	72		&	2	& 9.91	&	32	& 9.25 & 9.58	&	32	&	7.26		&	28$\%$		&	7.23 		&35$\%$	\\
NGC 1705 		&	4.7	&	8.46 	&	63	&	-	&	- 	&	114 $^a$	&     12	& 7.61	&	31	& -        &     -	&	-	&       4.46		& 	81$\%$		&	5.23 	$^b$	&28$\%$	\\	
NGC 3627		&	8.9	&	8.43 	&	44	&	-	&	- 	&	1860		&	11	& 8.88	&	13	& 9.52 & 9.92	&	14	&	7.92		& 	91$\%$		&	7.41 		&21$\%$	\\
Mrk 33			&	20	&	8.4 	&	57	&	-	&	- 	&	40		&	11	& 8.47	&	33	& 7.25 & 7.68	&	34	&	6.67		& 	15$\%$		&	6.51 		&12$\%$	\\
NGC 3190		&	17.4	&	8.4	&	14	&	-	&	-	&	190		&	11	& 8.65	& 	13	& -        &   -	&	-	&	7.38		& 	26$\%$		&	7.05 		&22$\%$	\\

\end{tabular}
 \end{table} 
 \end{landscape}



\addtocounter {table}{-1}

\begin{landscape}
\begin{table}[h!]
\caption[continued]{continued}
\label{Effects_of_the_submm_Table}
 \centering
 \begin{tabular}{p{1.8cm}p{1.2cm}p{0.8cm}p{0.5cm}p{0.8cm}p{0.5cm}p{1cm}p{0.5cm}p{0.8cm}p{0.8cm}p{1.2cm}p{1.2cm}p{0.8cm}p{1cm}p{1cm}p{1cm}p{1cm}}
\hline
\hline
(1) &  (2) &\multicolumn{2}{c} {(3)} & \multicolumn{2}{c} {(4)} &\multicolumn{2}{c} {(5)} & \multicolumn{2}{c} {(6)} &\multicolumn{3}{c} {(7)} &\multicolumn{2}{c} {(8)} & \multicolumn{2}{c} {(9)}\\
&&&&&&&&&&&&&&&\\
Name& Distance &\multicolumn{2}{c} {12+log(O/H)} &\multicolumn{2}{c} {S$_{450}$} &\multicolumn{2}{c} {S$_{850}$} & \multicolumn{2}{c} {log HI}&\multicolumn{3}{c} {log H$_2$} & \multicolumn{2}{c} {log M$_{dust~}$$_{w/o~submm}$} & \multicolumn{2}{c} {log M$_{dust~}$$_{with~ submm}$}\\
&&&&&&&&&&&&&&&&\\
 &           & Value & Ref. & Value & Ref. & Value & Ref. & Value &Ref. & {\revisedbis Value $_{Gal}$} & {\revisedbis Value $_{B02}$} & Ref. & Value   & unc                           & Value       & unc    \\
 &(Mpc) &            &          & 	(Jy)    &         &   (mJy) &        &(\msun)&        &  (\msun)    &(\msun)                &          & (\msun)& $\%$~M$_{dust}$ &  (\msun)   &    $\%$~M$_{dust}$ \\
\hline
&&&&&&&&&&&&&&&&\\

NGC 7331		&	14.7  & 	8.36 	&	44	 &	20.6	&	11	&	2110		&	11	& 9.96	&	14	& 9.90 & 10.37	&	14	&	8.03		& 	43$\%$		&	7.97 		&13$\%$	\\
NGC 3521		&	9.0	& 	8.36  &	44	&	-	&	- 	&	2110		&	11	& 9.75	&	13	& 9.55 & 10.02	&	14	&	7.67		& 	20$\%$		&	7.54 		&23$\%$	\\
NGC 4670		&	11	&	8.3 	&	17	&	-	&	- 	&	49     	&	5	& 8.22	&	5	& 7.01 & 7.55	&	5	&	5.90		& 	32$\%$		&	6.02 		&21$\%$	\\
NGC 2976		&	3.6	&	8.3	&	14	&	-	&	-	&	610		&	11	& 8.12	&	14	& 7.84 & 8.38	&	14	&	6.27		& 	38$\%$		&	6.18 		&35$\%$ \\
NGC 5253 		&	3.2	&	8.2	&	58	&	-	&	- 	&	180		&	7	& 7.96	&	35	& 6.91 & 7.55	&	36	&	6.15		& 	69$\%$		&	$<$ 5.95 	& -\\
NGC 1569 		&	2.2	&	8.2   &	52	&	1.33&	6 	&	280		&	6	& 8.11	&	37	& 5.86 & 6.50	&	38	&	5.24		& 	64$\%$ 		&	5.02 	$^b$	& 75 $\%$ \\	
NGC 5929 		&	33	&	8.18	&	5	&	-	&	- 	&	1190		&	5	& 8.63	&	5	& 8.07 & 8.73	&       5	&	7.45		& 	11$\%$		&	$<$ 7.34	& -\\	
NGC 1482		&	22	&	8.12	&	14	&	-	&	-	&	330		&	11	& 8.88	&	13	& 9.53 & 10.25	&	14	&	8.03		& 	44$\%$		&	7.40 		&39$\%$ \\
 II Zw 40 			&	10	&	8.1	&	59	&	0.24	&	8 	&	87		&	8	& 8.62	&	22	& 7.50 & 8.24	&	22	&	 5.56		& 	 56$\%$		&	 5.75 	$^b$	& 44$\%$ 	\\
UM 448			&	70	&	8.0   &	60	&	-	&	- 	&	39		&	5	& 9.67	&	39	& 9.38 & 10.22	&	39	&	7.23		& 	39$\%$		&	7.54 		& 51$\%$	\\
Mrk 1089			&	59.8&	8.0	&	62	&	-	&	- 	&	67.4$^a$	&     12	& 10.43	&	40	& -        &  -	&        - 	&	7.24		& 	31$\%$		&	7.72 		&20$\%$	\\
NGC 1140  		&	25	&	8.0   &	58	&	0.27	&	8 	&	69		&	8	& 9.71	&	41	& -        &  -	&	-	&	 6.61		& 	 29$\%$		&	 6.52 	$^b$	& 37$\%$ 	\\
Haro 11 			&	92	&	7.9   &	42	&	-	&	- 	&    40 $^a$ 	&	  12	& 8.0 	&	42	&$<$9.0&$<$9.0&	42	&	6.90		& 	45$\%$		&	7.3 	$^b$	&17$\%$ 	\\
I ZW 18			&	13	&	7.2	&	61	&	-	&	- 	&	$<$ 1.25	&	  7	& 8.08	&	43	& -        & -		&	- 	&	4.57		& 	61$\%$		&	$<$ 4.74	& -\\	
\hline
\end{tabular}
\begin{list}{}{}
\footnotesize
\item[Notes.-] (1) Galaxy Name. (2) Distance to object (References on distances are given in \citet{James2002}, \citet{Draine2007}, \citet{Galliano_Dwek_Chanial_2008} and \citet{Galametz2009}). (3) Metallicity of the galaxy.   (4) 450 \mic\ flux. (5) 850 \mic\ flux [ $^a$ 870 \mic\ fluxes estimated through \lab\ observations ].  (6) H{\sc i} mass of the galaxy. (7) H$_2$ mass of the galaxy. (8) Total dust mass estimated from our SED modelling performed with measures below 300 \mic. (9) Total dust mass estimated from our SED modelling performed with submm data [ $^b$ A cold dust component of 10K with an emissivity $\beta$=1 is added to the model to account for the submm excess ].
\item[References.-]  [1] \citet{Smith1990} - [2] \citet{Dunne2000} - [3] \citet{Dunne_Eales_2001} - [4] \citet{Klaas2001} - [5] \citet{James2002} - [6] \citet{Galliano2003} - [7] \citet{Hunt2005} - [8] \citet{Galliano2005} - [9] \citet{Stevens2005} - [10] ~\citet{Meijerink2005} - [11] \citet{Dale2007} - [12] \citet{Galametz2009} - [13] ~\citet{Kennicutt2003} - [14] ~\citet{Draine2007} - [15] ~\citet{Tilanus_Allen_1993} - [16] ~\citet{Lundgren2004} - [17] ~\citet{Lisenfeld_Ferrara_1998} - [18]~\citet{Dahlem2001} - [19]~\citet{Dahlem1990} - [20]~\citet{Bernloehr1993} - [21]~\citet{Yun2001} - [22]~\citet{Bettoni2003} - [23]~\citet{Ondrechen1989} - [24]~\citet{Gerin1988} - [25]~\citet{Appleton1981} - [26]~\citet{Walter2002} - [27]~\citet{Staveley1987} - [28]~\citet{Helfer2003} -  [29]~\citet{Sauvage1997} -  [30]~\citet{Dean1975} -  [31]~\citet{Meurer1998} - [32]~\citet{Struck2003} - [33]~\citet{Thuan2004} - [34]~\citet{Israel2005} - [35]~\citet{Reif1982} - [36]~\citet{Meier2002} - [37]~\citet{Stil_Israel_2002} - [38]~\citet{Israel1997} - [39]~\citet{Sage1992} - [40]~\citet{Williams1991} - [41]~\citet{Hunter1994} - [42]~\citet{Bergvall2000} - [43]~\citet{VanZee1998} - [44]~\citet{Walter2008} - [45]~\citet{Otte2001} - [46]~\citet{Pilyugin2004} - [47]~\citet{Webster1983} - [48]~\citet{Ravindranath2001} - [49]~\citet{Storchi-Bergmann_1995} - [50]~\citet{Boselli2002} - [51]~\citet{Dutil1999} - [52]~\citet{Kobulnicky_Skillman_1997} - [53]~\citet{Bonatto1990} - [54]~\citet{Bresolin2004} - [55]~\citet{Petrosian1993} - [56]~\citet{Gonzalez-Delgado_1995} - [57]~\citet{Mas-Hesse_1999} - [58]~\citet{Heckman1998} - [59]~\citet{Perez-Montero_2003}  - [60]~\citet{Izotov_Thuan_1998}  - [61]~\citet{Izotov1999} - [62]~\citet{Hopkins2002} - [63]~\citet{Meurer1992} - [64] http://sings.stsci.edu/Sample/
 \end{list}
 \end{table} 
 \end{landscape}


The SEDs of the 9 galaxies showing a submm excess and modelled with an extra cold dust component are presented in Fig.~\ref{SED_CD}. {\revised We already want to note that those galaxies only constitute a small subset of our sample and global relations studied in the following should not be affected by those objects. Those galaxies are marked by red symbols in the following plots.}

\begin{figure}
    \centering
    \begin{tabular}{p{0mm} m{8cm}}
     a) &  \includegraphics[width=8cm ,height=7.2cm]{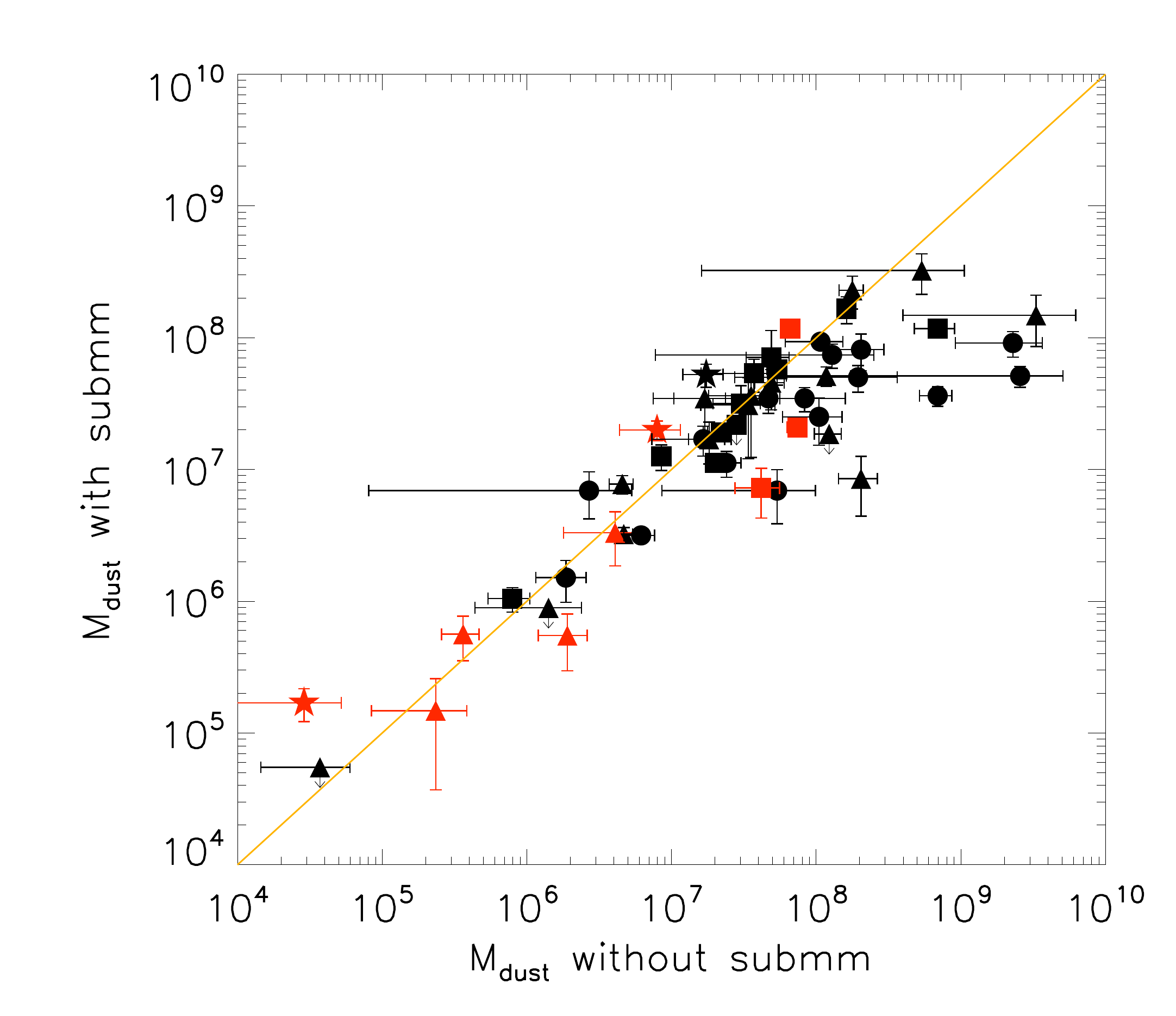} \\
     b) &  \includegraphics[width=8cm ,height=7.2cm]{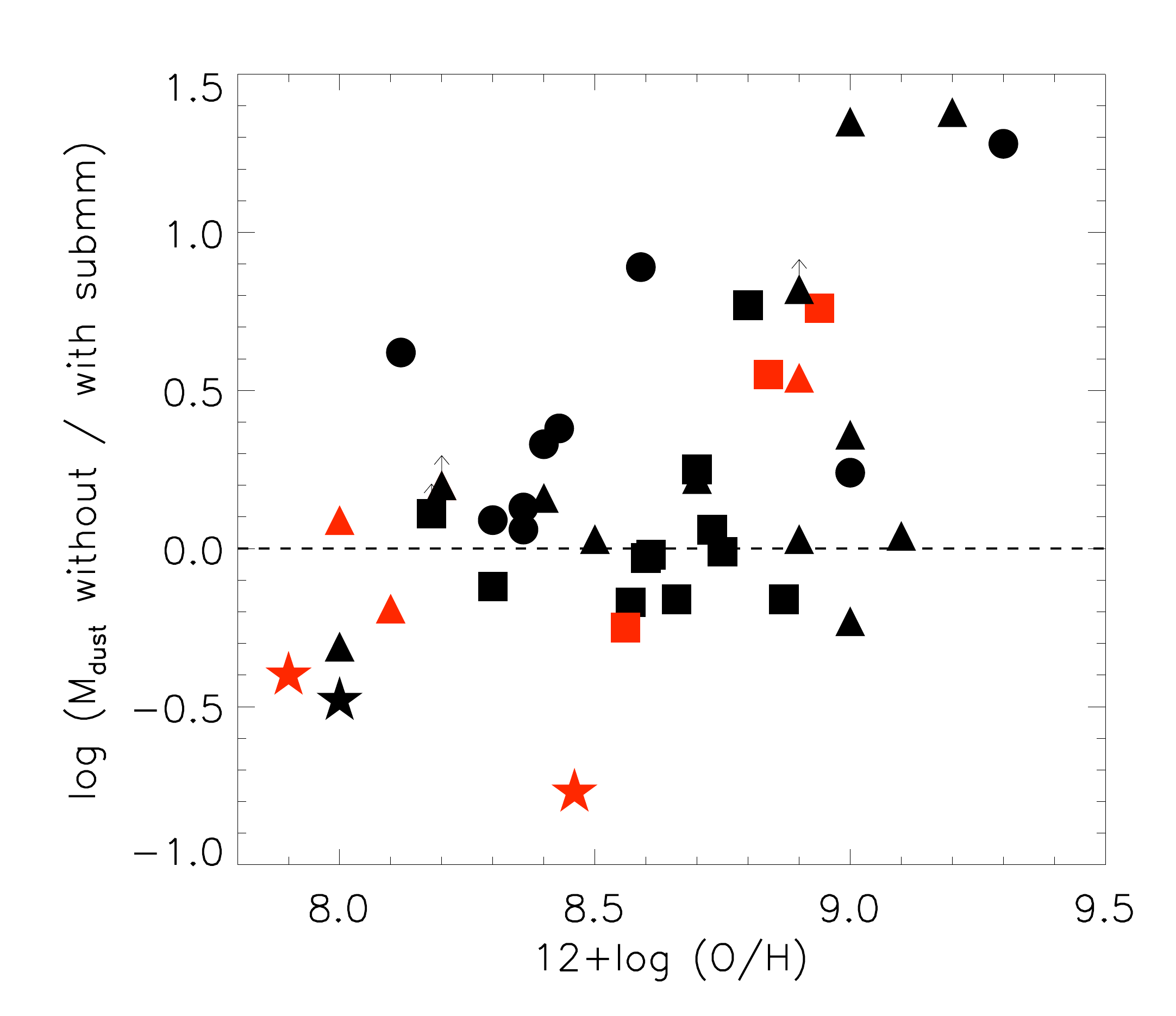} \\
       \end{tabular}
    \caption{ {\it a)} Total dust mass estimated with submm observations versus total mass of dust estimated without submm observations. The line shows the 1:1 relation. Squares indicate the \citet{James2002} galaxies. Circles indicate the galaxies of the SINGS survey \citep{Draine2007} for which metallicity is already published. Triangles are the \citet{Galliano_Dwek_Chanial_2008} sample and stars indicate the location of the three galaxies Haro~11, NGC~1705 and Mrk~1089 of ~\citet{Galametz2009}. Red symbols show the galaxies that present a submm excess and for which a cold dust (10K) component with an emissivity $\beta$=1 is added. {\it b)} ~Evolution of the ratio of the total mass of dust estimated with and without submm data as a function of the oxygen abundance of the galaxy.}
    \label{Effects_of_the_submm_Figure}
\end{figure}

\section {Analysis}

\subsection{Dust mass estimates with submm observations}

We apply our SED modelling technique to the observational constraints we gathered. We derive dust mass estimates with all the data including the submm data, and another time omitting the submm data, which normally means that the observations stop at 160 \mic. To quantify the errors on our dust masses for each SED solution, we produce a grid of 100 randomly modified observational constraints, allowing the observational fluxes to vary within their error bars, following a Gaussian distribution around their reference value. Dust masses and errors are summarized in the last column of Table~\ref{Effects_of_the_submm_Table}. 

We also test the presence of the mid-IR spectroscopic observations (e.g. ISO or \spitz\ IRS) on the dust mass deduced without submm, to study the influence of submm data on the submm slope. We run our model with and without the mid-IR spectrum and without data beyond 200 \mic. We find no strong differences nor systematic increase or decrease of the total dust mass deduced when a mid-IR spectrum is added to constrain the SED, as long as there are at least a few observations at mid-IR wavelengths (i.e. IRAC and MIPS 24 \mic\ or IRAS 25 \mic).

In Fig.~\ref{Effects_of_the_submm_Figure}a, we show the values of the dust mass estimated with submm data vs the dust mass estimated without submm data. Figure~\ref{Effects_of_the_submm_Figure}b presents the ratio of these two quantities as a function of metallicity. Both figures demonstrate that, for dusty massive galaxies, {\it the dust mass tends to be higher when submm measures are omitted}. We also note that errors on the dust mass of galaxies (as given in Table~\ref{Effects_of_the_submm_Table}) can be significant when submm data are omitted, especially for dustier galaxies.\\

\begin{figure}
    \centering
    \begin{tabular}{c}
      \includegraphics[width=9cm ,height=6cm]{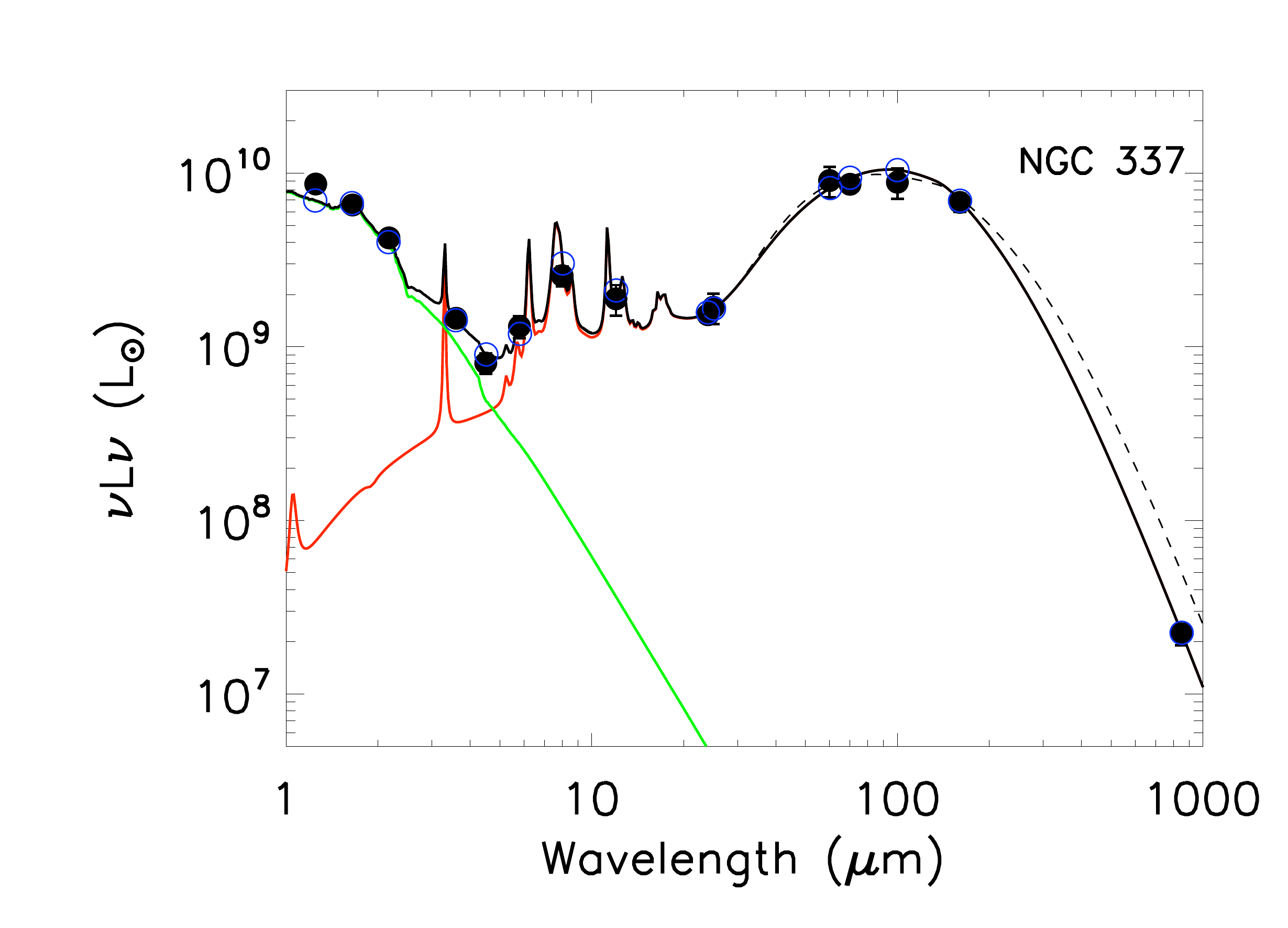} \\
       \end{tabular}
    \caption{Total SED of the galaxy NGC~337 using the 850 \mic\ SCUBA flux (solid line) and not using the submm information (dashed line). Observational constraints are superimposed (filled circles). Open circles are the expected modelled fluxes integrated over the instrumental bands. Green and red lines respectively distinguish the stellar and the dust contributions of the SED. }
    \label{NGC337_SED}
\end{figure}

In the case when submm data is lacking, there is indeed a risk that the SED procedure invokes a large mass of cool dust. Fig.~\ref{NGC337_SED} illustrates this effect on the galaxy NGC~337 modelled with (solid line) and without (dashed line) the 850 \mic\ SCUBA flux. The submm slope of the SED is reasonably sampled when the submm flux is available and the total dust mass decreases by a factor of 8 for this galaxy. This systematic overestimate of the dust masses mostly affects metal-rich galaxies. Indeed, their SEDs usually peak at longer wavelengths compared to the low-metallicity galaxies. The 160 \mic\ flux is thus no longer sufficient to fully sample the dust SED peak and constrain the submm slope. \\

{\revised The SCUBA/LABOCA fluxes constrain the submm slope of our SEDs and more particularly the minimum intensity U$_{min}$. This parameter is directly linked to the cold dust mass (see Fig.~\ref{nulnu_U} and the influence of $<$U$>$ in the emission of silicate and graphite) and thus significantly influences the total dust mass derived. An overestimate of U$_{min}$ will systematically lead to an underestimate of the dust mass. 
Figure~\ref{Uminwith_Uminwithout} shows the evolution of  U$_{min}$ parameter if submm observations are used in the modelling. The galaxies showing a submm excess do not appear on this plot. We observe a tighter dispersion of this parameter when submm data are used in the modelling. The small values of this parameter when submm observations are omitted lead to the high dust mass we derive in some cases. A similar study on the U$_{min}$ parameter has been performed by \citet{Draine2007}. They found that the dust masses of their SINGS-SCUBA sample is rather unaffected by the use of submm constraints.  
For the same galaxies, we find, on the contrary, that the dust masses we derive without using submm data can be significantly {\it higher} than those obtained using submm data.
Our model assumes a power-law distribution of the dust mass over interstellar radiation heating intensities. The SED model used in \citet{Draine2007} includes both this parametrisation and a ``diffuse" dust component heated by a single radiation field with an intensity of U=U$_{min}$ which describes most of the far-IR~/~submm emission of their galaxies. 
They also chose to restrict their radiation fields to intensities U$_{min}$ $\ge$ 0.7 as an {\it ad hoc} measure to avoid their model invoking a large mass of cold dust heated by weak starlight when submm observations are not available. 
The boundaries of the minimum heating intensity range chosen by \citet{Draine2007} are shown in the shaded region in Figure~\ref{Uminwith_Uminwithout}. The restriction of the U$_{min}$ interval seems to be consistent with their objects, and for most of the metal-rich galaxies of our sample. \\


\begin{figure}
   \centering
    \begin{tabular}{c }
       \includegraphics[width=8cm ,height=6.8cm]{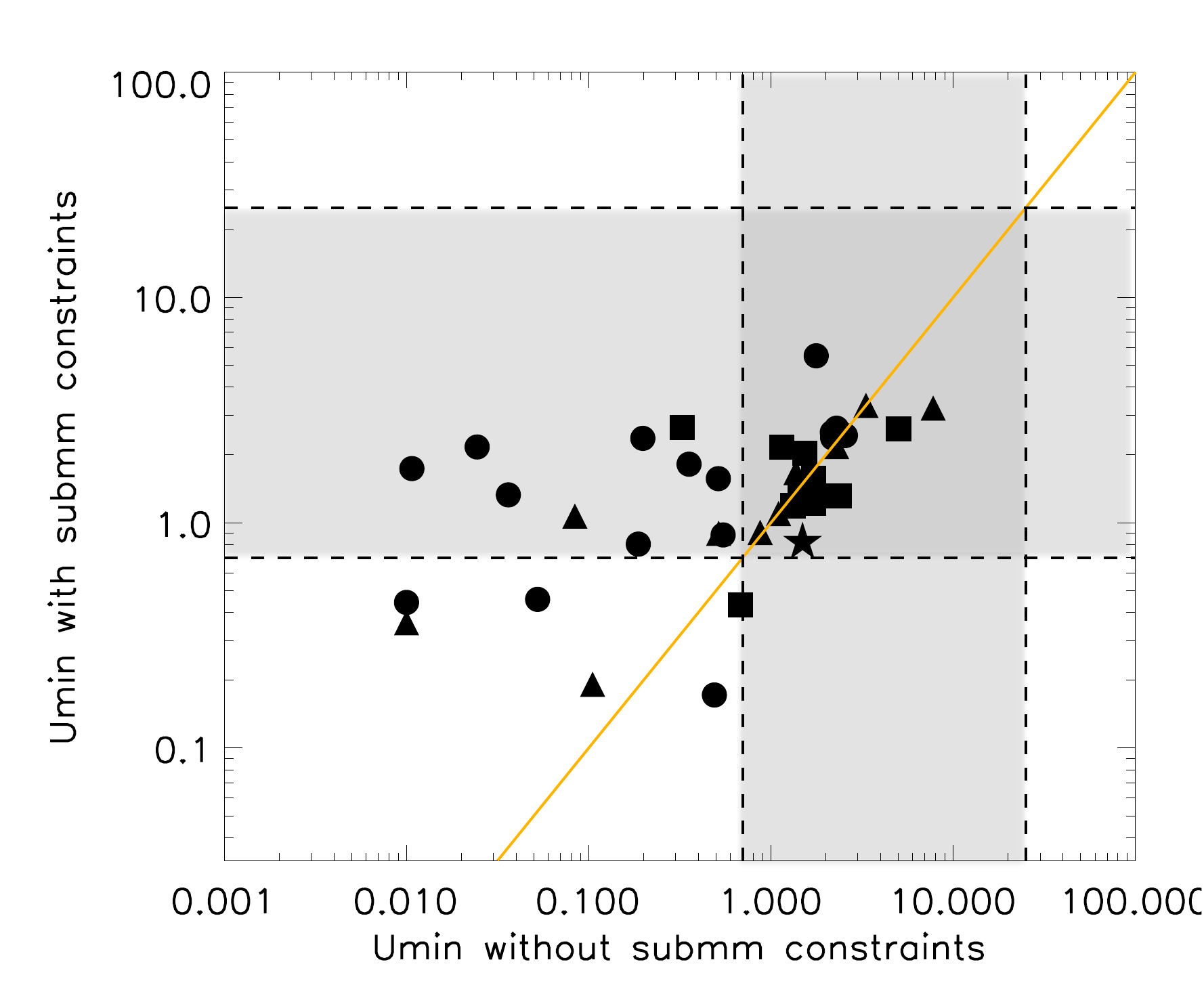}\\
         \end{tabular}
    \caption{ Minimum heating intensity derived from the SED model using submm data compared to minimum heating intensity determined without submm data. The orange line indicates the 1:1 relation. The dashed lines define the limits \citet{Draine2007} used as boundaries for their minimum heating intensity U$_{min}$. We remind the reader that our model solutions are obtained with U$_{min}$ as a free parameter. For convention on symbols, see Fig.~\ref{Effects_of_the_submm_Figure}. Limiting U$_{min}$ to 0.7 avoids an overestimate of the dust mass but removes the possibility of low temperature radiation fields.}
    \label{Uminwith_Uminwithout}
\end{figure}

However, as noted in the \citet{Draine2007} study, the SINGS-SCUBA galaxy sample could be biased in favor of increased dust mass and associated star formation. The restriction on the U$_{min}$ parameter would not be valid in systems containing a large amount of cold dust heated by weaker stellar light. This assumption could lead, in a systematic way, to an underestimate of the dust mass of those environements when submm data are not available and should be used with caution in some galaxies. 

Indeed, we find that for some low-metallicity galaxies of our sample, the dust masses estimated without submm observations are, on the contrary, {\it lower} than those determined with submm data (Fig.~\ref{Effects_of_the_submm_Figure}b). This increase of the total dust mass while using submm data was already discussed in \citet{James2002}. They observed some of the galaxies presented in \citet{Lisenfeld2001} with SCUBA and noticed an increase of the total dust mass of their galaxies when submm data are used. Indeed, the \citet{Lisenfeld2001} study was performed based on \iras\ 60/100 \mic\ fluxes only and could have missed cold dust at longer wavelengths and thus underestimated the total dust mass. For most of the low-metallicity galaxies of our sample, the restriction of the U$_{min}$ parameter when submm data are not available is an unsatisfactory solution, since such a bias leads to an {\it underestimate} of the total dust mass. In those environments, submm data are crucial to trace the coldest phases of dust and also detect, if present, a possible submm excess.

In conclusion, when comparing the results unconstrained to those with the proper submm data, we find that the uncertainty on the dust mass is non-trivial and systematic, depending on metallicity of the object. Without submm constraints, our fit of the SEDs will give higher mass for metal-rich galaxies and lower masses for low-metallicity galaxies. 

In Fig.~\ref{Effects_of_the_submm_Figure}, we finally note that our modelling of the submm excess (which includes a modified blackbody with T=10K - red symbols) does not systematically lead to an increase of the dust mass estimate of a galaxy compared to that estimated omitting submm data . The slope of the trend (i.e. the correlation coefficients) observed in Figure~\ref{Effects_of_the_submm_Figure}b is, furthermore, not affected by the addition of galaxies in which a submm excess is detected.}

\subsection{The Dust-to-Gas mass ratio relation with metallicity}

\subsubsection{The gas masses and metallicities}
 
Values and references of the gas masses of our galaxies are given in Table~\ref{Effects_of_the_submm_Table}. Both atomic and molecular gas masses are given when available in the literature and gas masses were evaluated using the same distances as those used to derive the dust masses and quoted in Table~\ref{Effects_of_the_submm_Table}. {\revisedbis We determine the molecular gas masses using the canonical galactic X$_{CO}$ factor of 2.3 $\times$ 10$^{20}$ cm$^{-2}$~/~K~km~s$^{-1}$ of \citet{Strong1988} (Table~\ref{Effects_of_the_submm_Table}, Value $_{Gal}$). The conversion factor from CO to H$_2$ (X$_{CO}$) in galaxies depends on the metallicity \citep{Wilson1995,Israel1997,Israel2000,Barone2000,Boselli2002,Strong2004, Leroy2005}. \citet{Boselli2002} studied X$_{CO}$ in normal late-type galaxies and derived X$_{CO}$ factors ranging from $~$10$^{20}$ cm$^{-2}$/K km s$^{-1}$ in giant spirals to 10$^{21}$ cm$^{-2}$/K km s$^{-1}$ in dwarf irregulars. They conclude that using a constant X$_{CO}$ factor to derive the molecular gas mass from CO observations could lead to an underestimation of a factor $\sim$10.
 We use the relation from \citet{Boselli2002} to estimate X$_{CO}$ for each galaxy depending on its metallicity and rescale the molecular gas masses (Table~\ref{Effects_of_the_submm_Table}, Value $_{B02}$):
 
 \begin{equation}
 log X_{CO} = -1.01 (12 + log(O/H)) + 29.28
 \end{equation}}


The metallicities of the galaxies are given in Table~\ref{Effects_of_the_submm_Table} and were estimated in the different papers using standard R$_{23}$ methods \footnote{The oxygen abundance, commonly used to estimate the metallicity, can be derived from the bright line observations and depends on the quantity [O{\sc ii}]$\lambda$3727+[O{\sc iii}]$\lambda$4959,$\lambda$5007 / H$\beta$, also called the R$_{23}$ parameter \citep{Pagel1979}.}, except in some cases discussed in $\S$ 4.2.2. We would like to warn the reader about the current debate on the correct way to estimate the metallicity of a galaxy. For instance, \citet{Pilyugin2005} discuss the problem of empirical calibrations performed with strong oxygen line intensities-oxygen abundances and propose revised methods to estimate the metallicity \citep[see][for applications on the \spitz/SINGS sample]{Moustakas2010}. These different methods will not affect the general trend of our sample but could, nevertheless, introduce a systematic offset in our metallicity values. 

\subsubsection{A revised D/G relation}

\begin{figure*}

    \centering
    \begin{tabular}{ p{6mm} m{13cm} }
    	 a) & \includegraphics[width=10.5cm ,height=7.5cm]{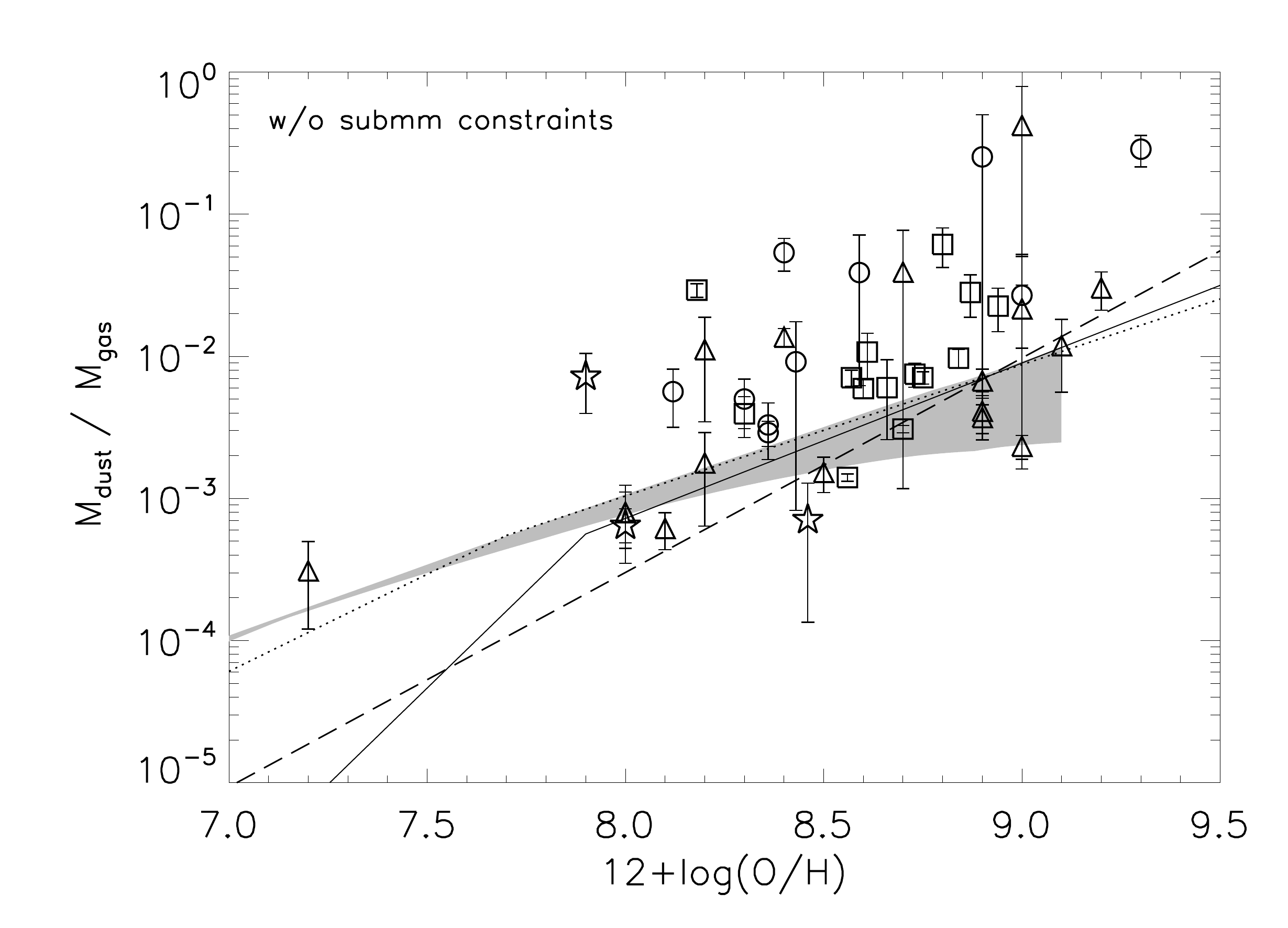} \\
	 b) & \includegraphics[width= 10.5cm ,height=7.5cm]{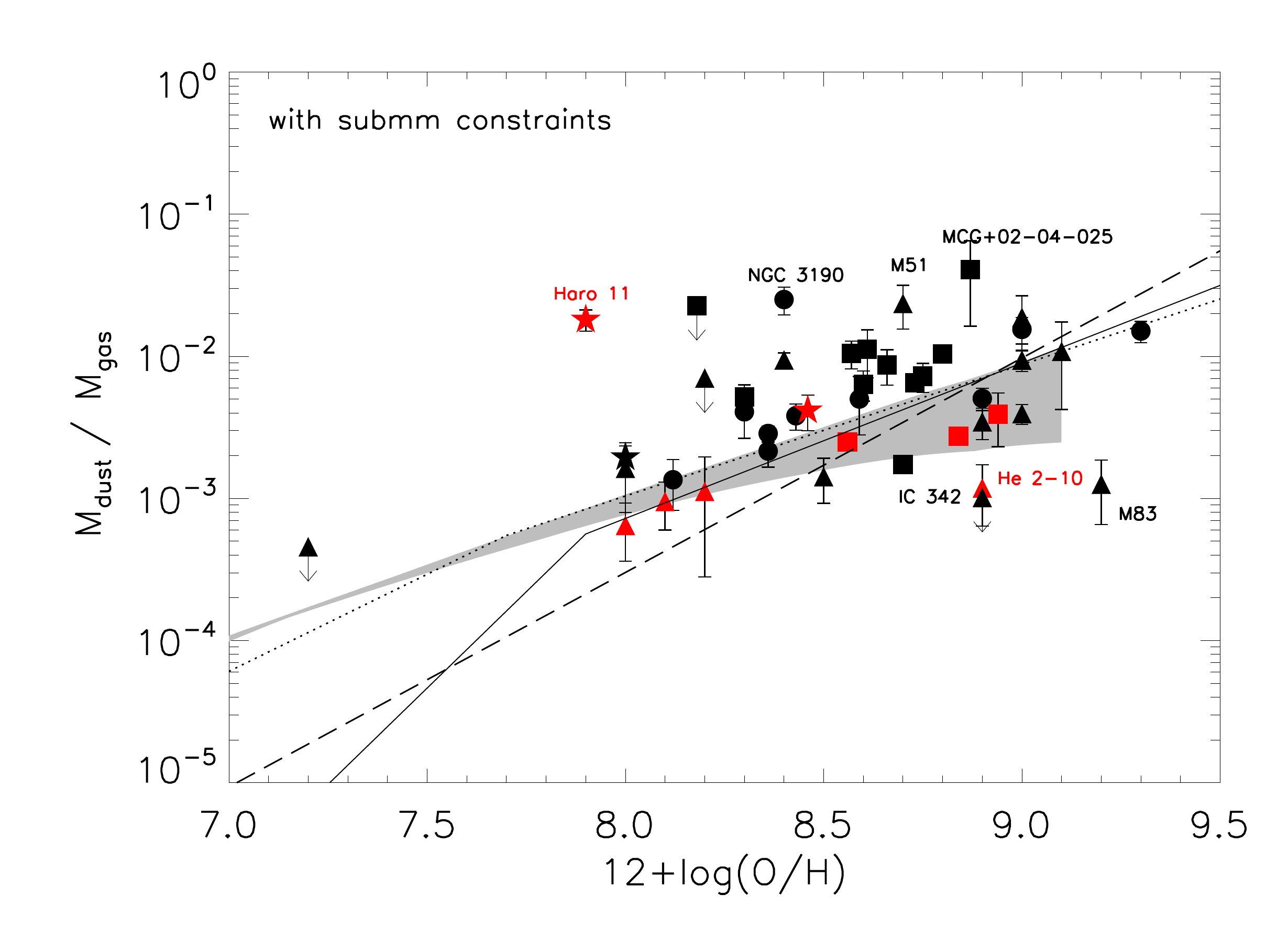} \\
    	 c) & \includegraphics[width= 10.5cm ,height=7.5cm]{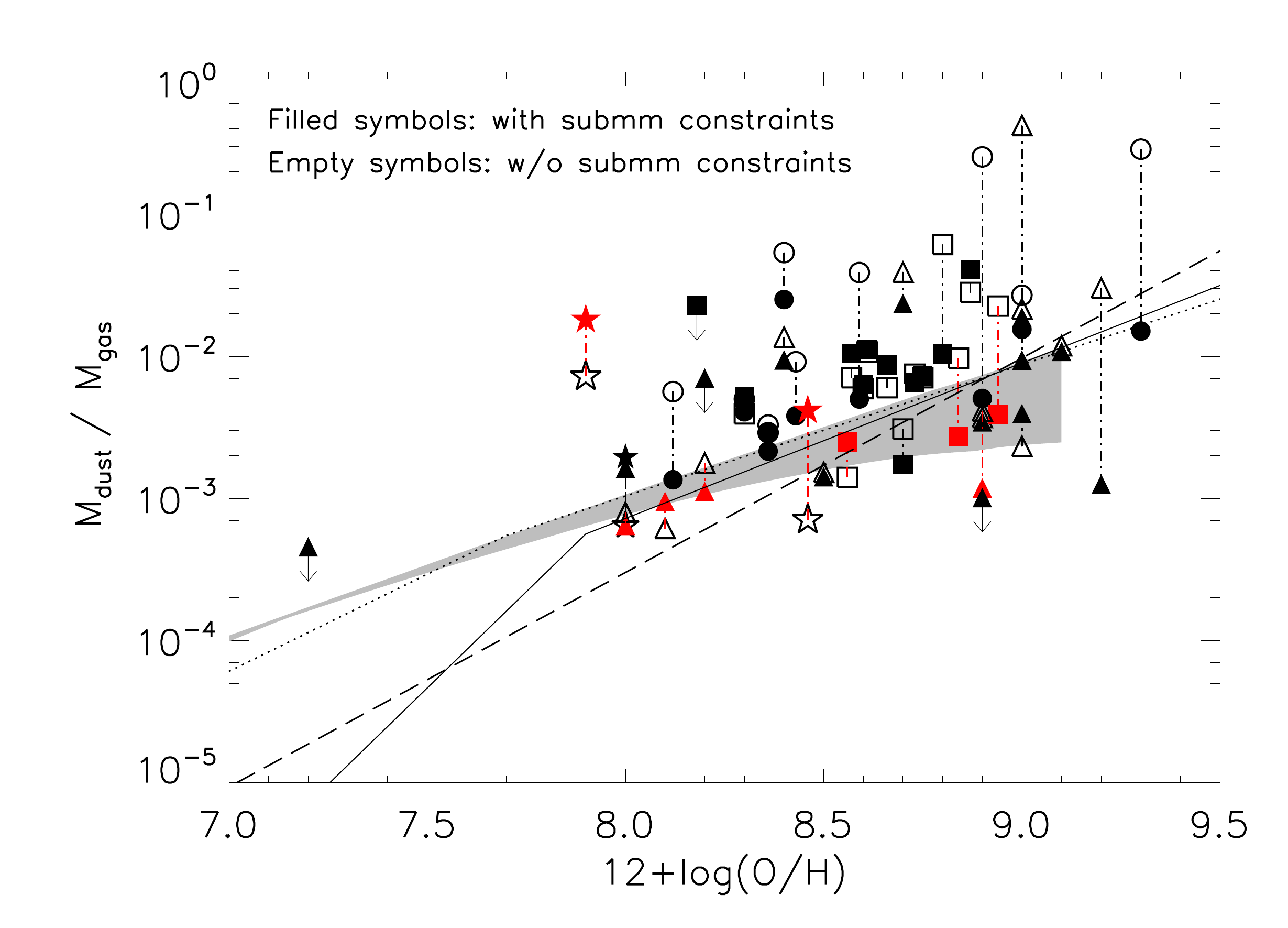} \\
     \end{tabular}
  \caption{{\it a)} Dust-to-gas mass ratio as a function of oxygen abundance for our restricted sample of galaxies that were observed at submm wavelengths. For the convention on symbols, refer to Fig.~\ref{Effects_of_the_submm_Figure}. Red symbols show the galaxies that present a submm excess and for which a cold dust (10K) component with an emissivity $\beta$=1 is added. In this plot, the dust masses of the galaxies are derived from our SED models without introducing submm observational constraints. The solid line and the dotted line are the predictions of the dust formation models from \citet{Edmunds2001}. The gray stripe is the expectation of the one-zone, single-phase chemical evolution model of \citet{Dwek1998} presented in \citet{Galliano_Dwek_Chanial_2008}. Both models are presented in $\S$ 2. The dashed line is the linear regression of the broad sample of $\S2$ where all galaxies, including those without submm data, are used. 
 {\it b)} Same but the dust masses of the galaxies are derived from our SED models using submm observational data. Labeled galaxies are discussed in $\S$ 4.1. Error bars refer to errors in the dust mass estimates and do not take the gas mass errors into account. {\it c)} We gather the two previous plots to emphasize how the D/G ``moves"  (dot-dash vertical lines) in the diagram when submm data are used to estimate the dust mass in the SED modelling.  
}
\label{Dust-to-gas_Metallicity_2}
\end{figure*}

Figure~\ref{Dust-to-gas_Metallicity_2} shows the D/G as a function of oxygen abundance. More precisely, the D/G is estimated using the dust masses calculated without submm fluxes in Fig.~\ref{Dust-to-gas_Metallicity_2}a and then recalculated using the submm data in Fig.~\ref{Dust-to-gas_Metallicity_2}b. The error bars indicate the errors on the dust masses. We separate the two plots for clarity. Figure~\ref{Dust-to-gas_Metallicity_2}c gathers information from Figure~\ref{Dust-to-gas_Metallicity_2}a and Figure~\ref{Dust-to-gas_Metallicity_2}b. The dot-dash vertical lines indicate how galaxies ``move" on the D/G axis when submm measures are used in the dust mass estimation and clearly show that for dustier galaxies, submm data significantly affect the total dust mass estimates by shifting the dust masses to lower values. The predictions of the \citet{Edmunds2001} and \citet{Dwek1998} models presented in $\S$ 2 are indicated on the plots. We also overlay the linear regression (dashed line) which was determined from our original broad sample of galaxies ($\S$ 2 and Figure~\ref{Dust-to-gas_Metallicity_1}).\\

Our sample restricted to galaxies observed at submm wavelength does not follow the regression of the original big sample presented in Figure~\ref{Dust-to-gas_Metallicity_1}. The D/G vs metallicity relation, in fact, now tends to flatten toward lower metallicity when submm data are used in the dust mass estimates, with D/G values of low-metallicity galaxies systematically higher using submm data than those determined without submm data. Furthermore, the use of submm data significantly tightens the D/G vs metallicity relation. {\revisedbis In order to quantify the scatter in both plots, we calculate the standard deviation of log(D/G) in the three metallicity bins Z=[8.2 , 8.5], Z=[8.5 , 8.8] and Z=[8.8 , 9.1]. The standard deviation is equal to 0.55, 0.44 and 0.72 in those different metallicity bins when submm data are omitted in the SED modelling to determine the total dust mass. Those values decrease to 0.34, 0.32 and 0.5 if submm data are used, consistent with the visual impression of a tightening in the D/G vs metallicity relation.}

We note that galaxies modeled with an extra cold dust component (red symbols) do not affect the global relation.
Unfortunately, the current observations are still not sufficient to discriminate between the two models of \citet{Edmunds2001}, mainly due to the lack of low-metallicity galaxies observed at submm wavelengths because of the difficulty to perform such observations. Submm observations from {\it Planck} and newly available ground based submm instruments (e.g. SCUBA-2, \lab) will be necessary to observe more low-metallicity environments at better sensitivity. Furthermore, the \hersc\ Guarantee Time Key Program, Dwarf Galaxy Survey (PI: S. Madden), is dedicated to the observation of a wide range of low-metallicity environments with the far-IR to submm PACS and SPIRE instruments. These observations will enable us to extend our knowledge on the D/G vs metallicity relation to lower metallicities, using 70 to 500 \mic\ data to constrain the peak and the submm slope of the SEDs.

 \subsubsection{Outlying galaxies} 

In this $\S$, we focus on the galaxies that deviate from the relation between the D/G and oxygen abundance.\\
 
{\it MCG+02-04-025} - The H$_2$ mass could not be estimated for the galaxy due to the lack of CO observations for this galaxy. The estimated D/G is thus an upper limit. \\

\begin{figure*}
    \centering
    \begin{tabular}{m{8.1cm} m{8.1cm}}
      \includegraphics[width=8.5cm ,height=5.5cm]{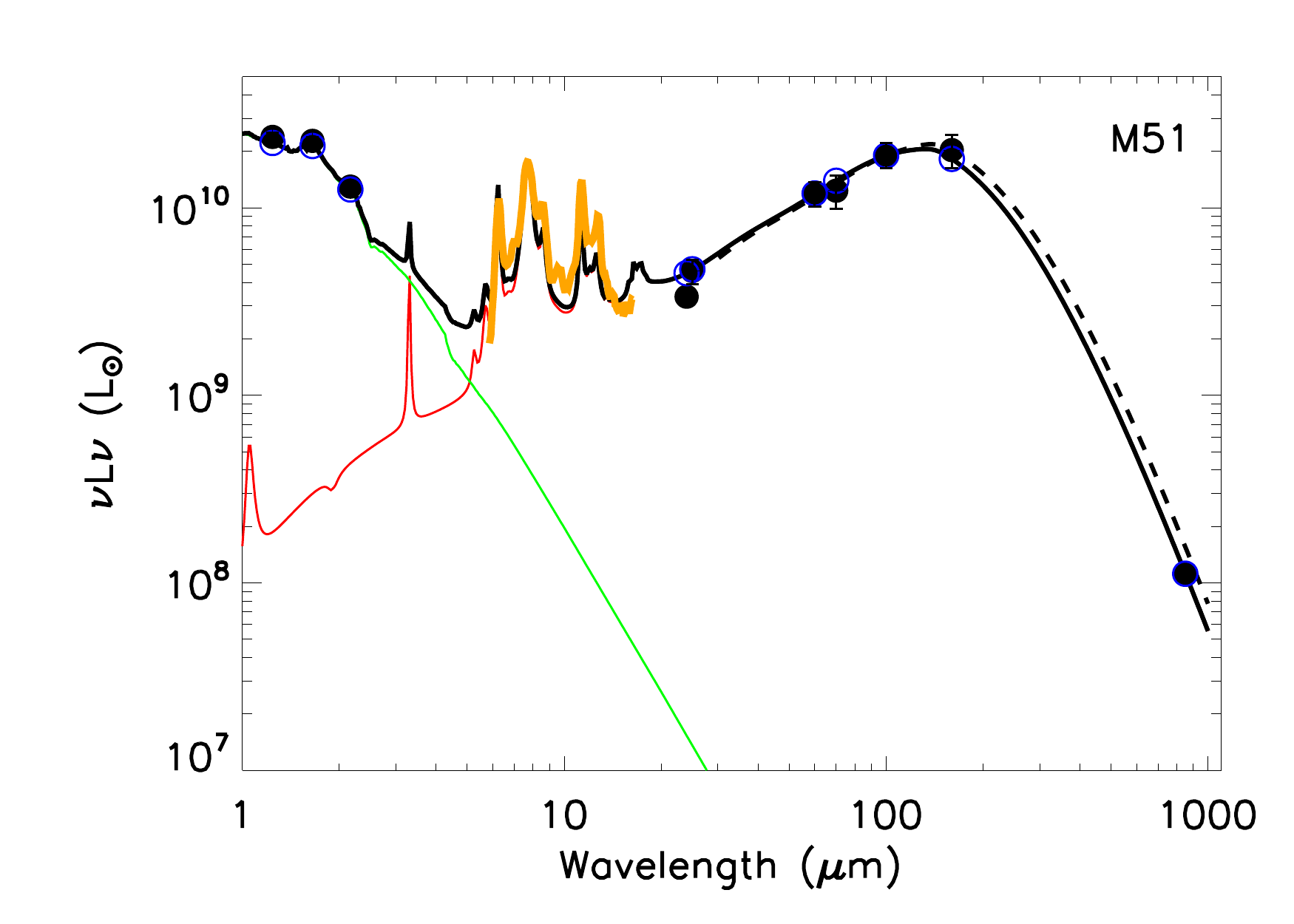}   &
       \includegraphics[width=8.5cm ,height=5.5cm]{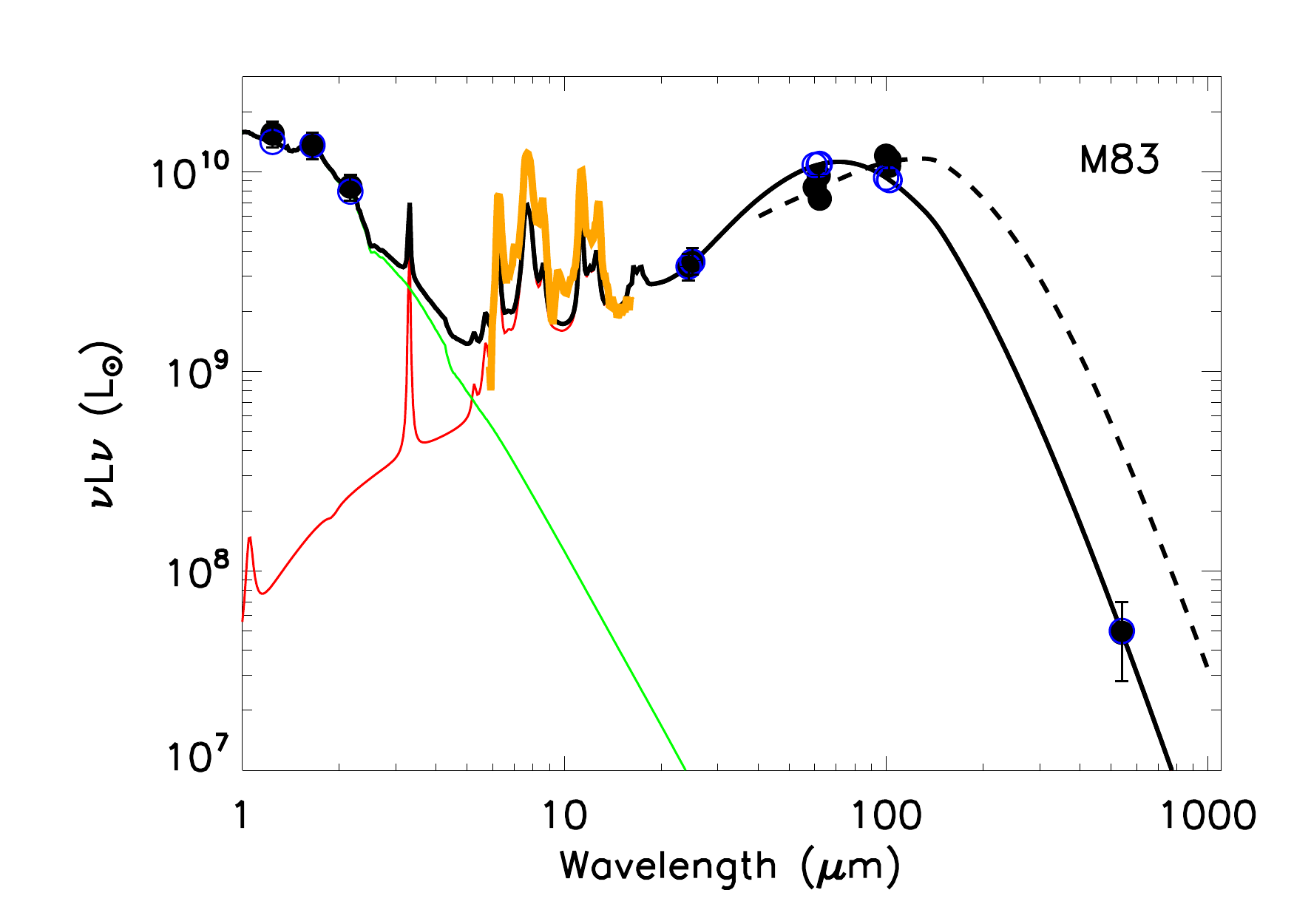}   \\
        \includegraphics[width=8.5cm ,height=5.5cm]{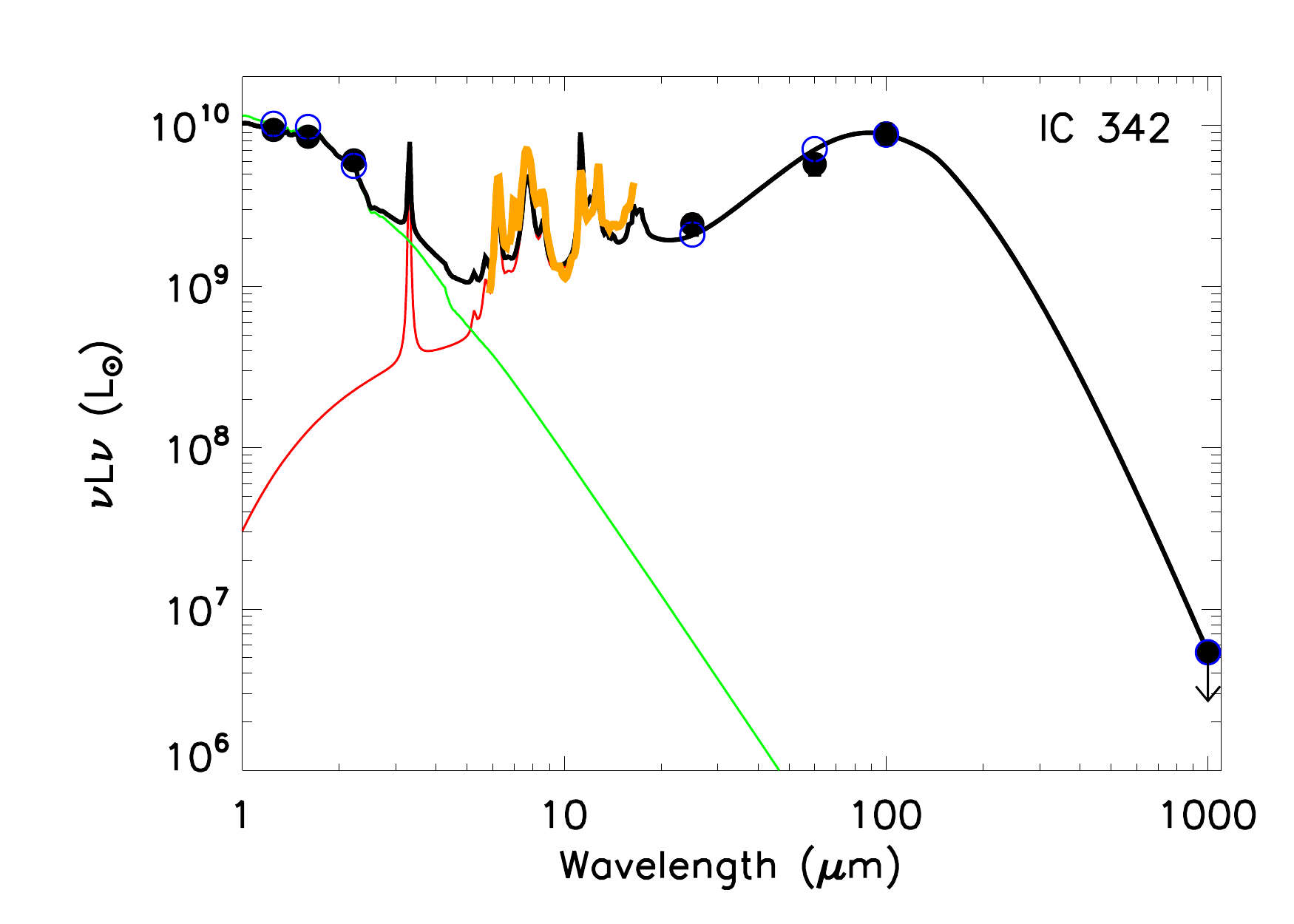}   &
      \includegraphics[width=8.5cm ,height=5.5cm]{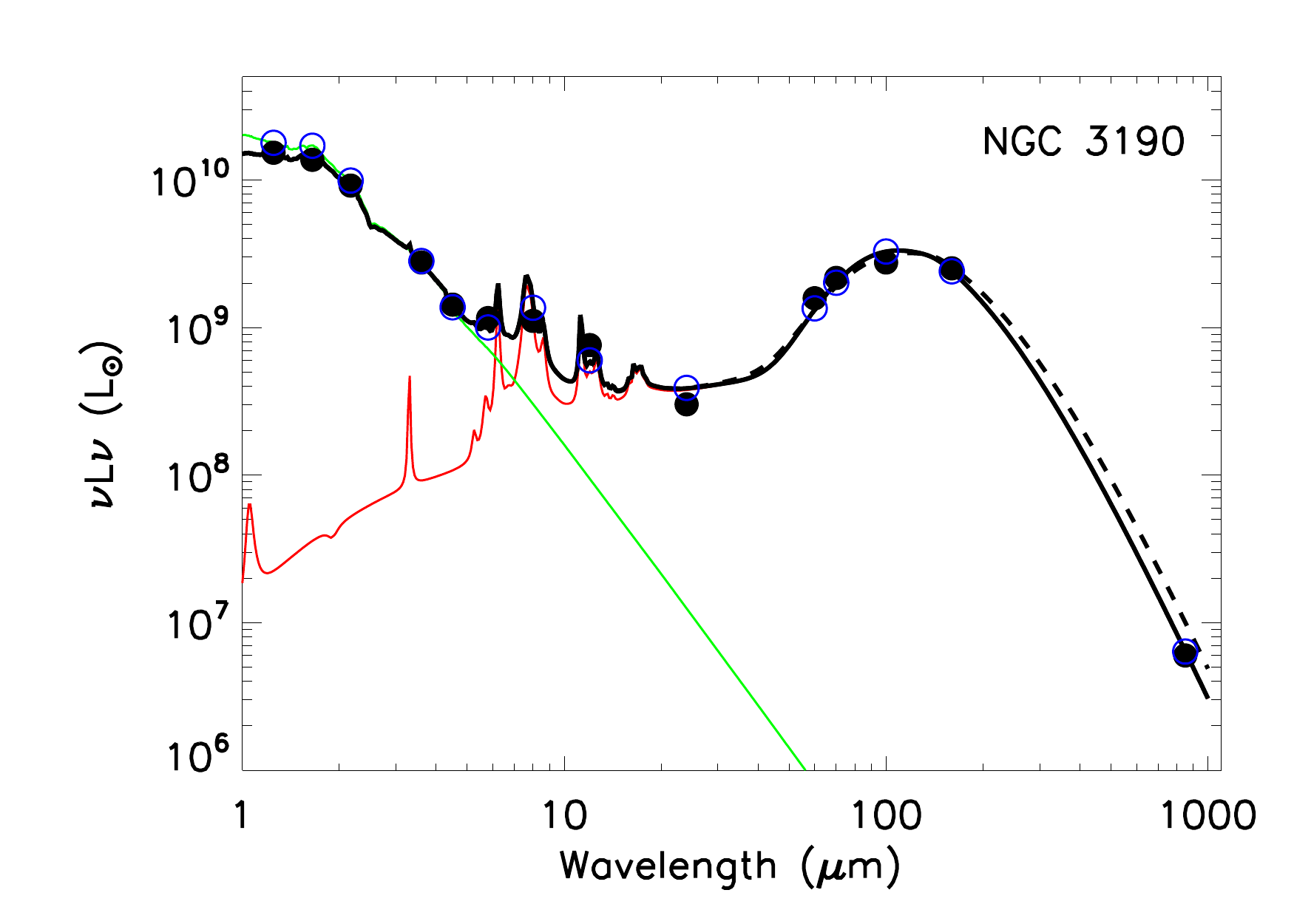} \\
       \end{tabular}
    \caption{SED models of the galaxies M51, M83, IC342 and NGC~3190. The SEDs modelled with submm data are indicated by the black lines. Dashed lines indicates the model obtained when submm data are omitted. Observational constraints are superimposed (filled circles). When the error bars are not shown, the errors are smaller than symbols. The thick orange line shows the IRS spectrum used in the SED modelling when available. The open circles indicate the expected modelled fluxes integrated over the instrumental bands. The green and red lines respectively distinguish the stellar and dust contributions.}
    \label{SED_others}
\end{figure*}

 {\it M51} - This galaxy has an unusually ``cold" global SED (Fig.~\ref{SED_others}). Applying only \iras\ data, the SED modelling gives much lower total mass of dust but the use of \spitz\ data, especially the 160 \mic\ of MIPS already significantly increases the dust mass, thus increases the D/G. There are still difficulties in estimating an average metallicity due to the extension of the galaxy and the fact that metallicity usually varies as a function of radius for spiral galaxies \citep[e.g.][]{Phillipps1991,Issa1990}. For M51, \citet{Moustakas2006} also estimated the metallicity to be 12+log(O/H)=8.7 using \citet{Pilyugin2005} but their estimate reaches 9.2 using the \citet{McGaugh1991} strong-line calibration of the R$_{23}$. A higher metallicity would, in fact, make the galaxy move rightward, close to the \citet{James2002} relationship but we prefer to keep the value 8.7 estimated by \citet{Bresolin2004}, which is based on high-quality H{\sc ii} region observations and calculated in the arms of the galaxy, where star-forming regions are concentrated. Nevertheless, M51 could have a strong gradient of metallicity that affects its position on this plot.\\
 

  {\it M83} - The SED of this galaxy is plotted in Fig.~\ref{SED_others}. This galaxy is surrounded by a very large H{\sc i} envelope, with a radius $>$ 125 kpc - 6.5 times the optical radius  \citep{Huchtmeier_Bohnenstengel_1981}. The total H{\sc i} mass of this envelope was estimated to be 2 $\times$ 10$^{10}$ \msun\ \citep{Tilanus_Allen_1993}. The H{\sc i} mass of Table~\ref{Effects_of_the_submm_Table} only accounts for that included in the photometric apertures used to estimate the fluxes we use in the SED modelling to derive the total mass of dust. The low D/G we derive is thus not due to an overestimation of the H{\sc i} mass. \citet{Lundgren2004} studied the kinematics of molecular gas of this galaxy with CO observations performed at the SEST and estimate the total gas mass, including H$_2$ and H{\sc i} to be $\sim$ 7.8 $\times$ 10$^9$\msun, close to that used by \citet{Galliano_Dwek_Chanial_2008} for the D/G estimate plotted in Fig.~\ref{Dust-to-gas_Metallicity_2}. The metallicity used in this paper (12+log(O/H)=9.2) was determined by ~\citet{Webster1983} using the R$_{23}$ method used in our study. \citet{Engelbracht2008} recently re-estimated the oxygen abundance of M83 with the direct electron-temperature (Te) method applied to individual H~{\sc ii} regions \citep{Skillman1998} and found 12+log(O/H)=8.62. This metallicity would thus bring the galaxy closer to the model predictions. \\
    
 {\it IC342} - The submm constraint we used is the 1mm flux and is an upper limit (Fig.~\ref{SED_others}), meaning that the dust mass we estimated is also an upper limit too. The location of IC342 at low Galactic latitudes makes it difficult to study because of its large extent. It is affected by foreground extinction due to our Galaxy and the  background subtraction issues it generates. This galaxy possesses a very peculiar optical/near-infrared shape \citep{Galliano_Dwek_Chanial_2008}. \\
 
  {\it He 2-10} - This Blue Compact Dwarf (BCD) differs from other BCDs because of its near solar metallicity \citep{Kobulnicky1999_2} and contains a relatively large amount of molecular gas  \citep{Meier2001}, which is unusual for low-metallicity dwarf galaxies. Nevertheless, the metallicity of the galaxy is rather uncertain due to the lack of measured [O III]  $\lambda$4363. \\ 
  
  {\it NGC~3190} - No H$_2$ mass estimates have been obtained in NGC~3190. The value presented on the plot is thus an upper limit. An amount of molecular gas equivalent to that of the H{\sc i} mass would lead to a higher D/G. The SED of this galaxy is plotted in Fig.~\ref{SED_others}  \\

{\it Haro~11} -  Using \spitz\ data, \citet{Engelbracht2008} note that Haro~11 possesses an intriguing SED, dominated by the mid-IR emssion and showing weak emission at 70 \mic. A D/G of $\sim$ 10$^{-3}$ would be expected, using the \citet{Galliano_Dwek_Chanial_2008} chemical model, for a galaxy with such metallicity (12+log(O/H)=7.9). We find an unusually high dust/H{\sc i} mass ratio $<$ 2 $\times$ 10$^{-1}$, using the H{\sc i} mass upper limit of 10$^8$\msun\ given by \citet{Bergvall2000}. This high value could suggest that the scenario of cold dust may be physically inappropriate for this galaxy if the totalgas reservoir is estimated correctly. In the particular case of Haro~11, \citet{Galametz2009} show that, even taking all the possible upper limits of gas into account in the total gas mass estimate (H$_{2}$ derived from CO(1-0) observations in \citet{Bergvall_Ostlin_2002} or from [CII] measurement in \citet{Bergvall2000}, ionized gas mass estimate in \citet{Ostlin1999}), the galaxy is an outlier of the Schmidt law~\citep{Kennicutt1998}, relation which relates the star formation rate of a galaxy to its total IR luminosity. They derived a ``missing gas mass" of $\sim$ 10$^{10}$ \msun, a value consistent with that independantly derived using the dust mass. A large amount of gas is thus not `detected' by current observations in Haro~11. \\

{\revised The molecular gas mass, commonly inferred from the CO tracer, is very difficult to estimate in low-metallicity galaxies. The CO to H$_2$ conversion factor is usually calibrated with cold ($\approx$ 12K) Galactic clouds (solar metallicity) and may not necessarily resemble the physical conditions of low-metallicity objects. Some studies show that this conversion factor can be far higher in low-metallicity environments than in metal-rich ones \citep{Wilson1995,Madden1997,Leroy2005,Leroy2007}. Due to the high excitation and the paucity of the dust in low-metallicity environments, CO is generally a poor tracer of molecular gas. The opacity of the H$_2$ and CO lines can create a condition where the molecules are self-shielded, an effect which can be efficient for hiding H$_2$ in regions where CO is photodissociated. Cold dust could reside in small shielded molecular/cold dust cores embedded in [CII] emitting envelopes where UV radiation would hardly penetrate. High values of L([CII])/L(CO) have indeed been found in low-metallicity galaxies sometimes up to 10 times higher than normal dusty starbursts or spirals \citep{Poglitsch1995, Madden1997, Madden2000, Hunter2001,Galliano2003}. This high ratio of observed L([CII])/L(CO) generally suggests a clumpy ISM. It is, for instance, the case in the galaxy Haro~11 discussed above, with L([CII])/L(CO)$\ge$ 10$^5$ \citep{Bergvall2000}. CO is also known to be a poor tracer of dense gas. Further observations of the far-IR fine structure lines will be done with the \hersc\ Dwarf Galaxy Survey (PI: S. Madden) to obtain a more accurate estimate of the gas content of dwarf galaxies.}

\subsection {Conditions for submm excess}

\begin{figure}[h!]
   \centering
    \begin{tabular}{p{0mm} m{8cm} }
       a) & \includegraphics[width=8cm ,height=6.8cm]{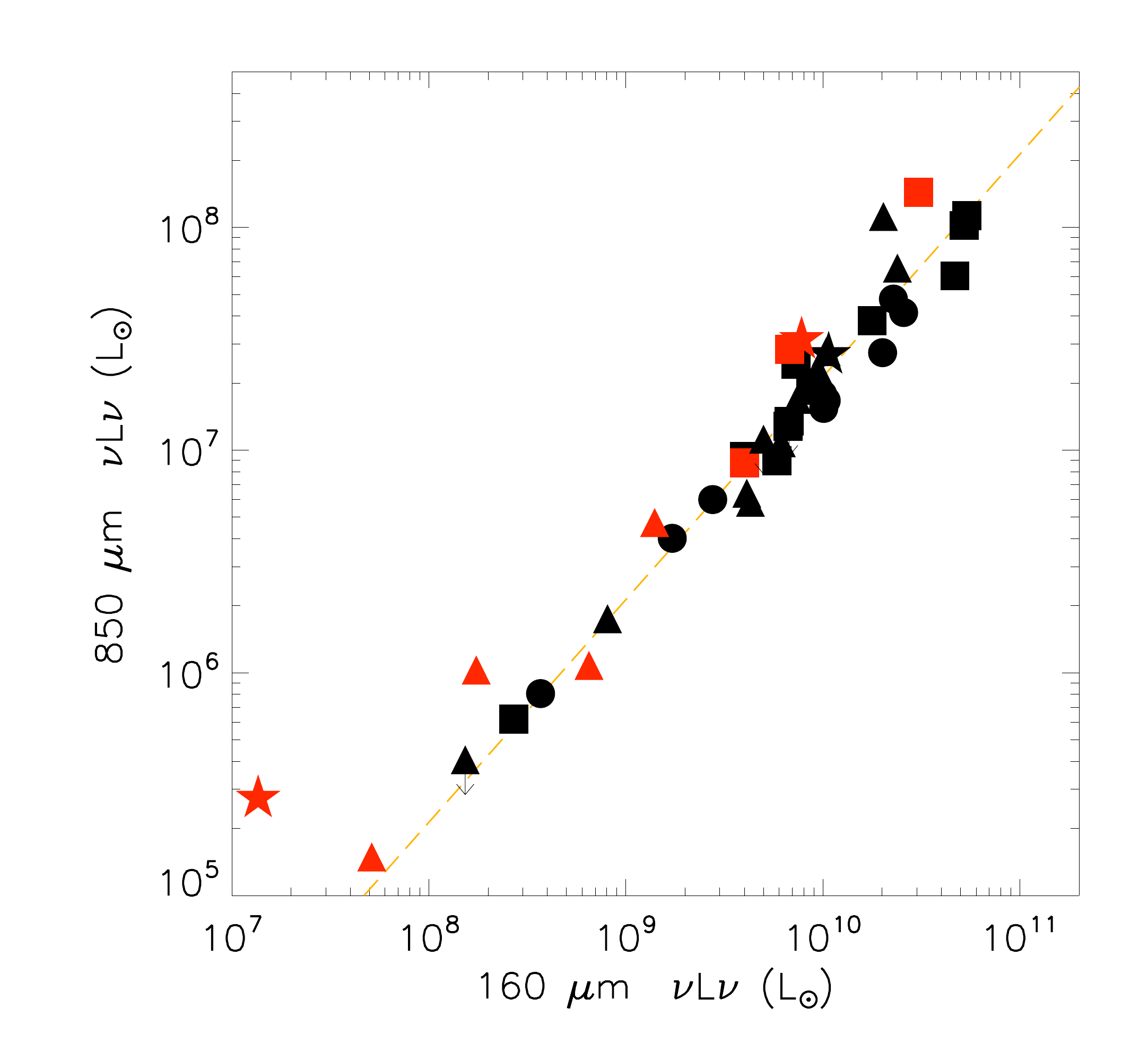}  \\
       b) & \includegraphics[width=8cm ,height=6.8cm]{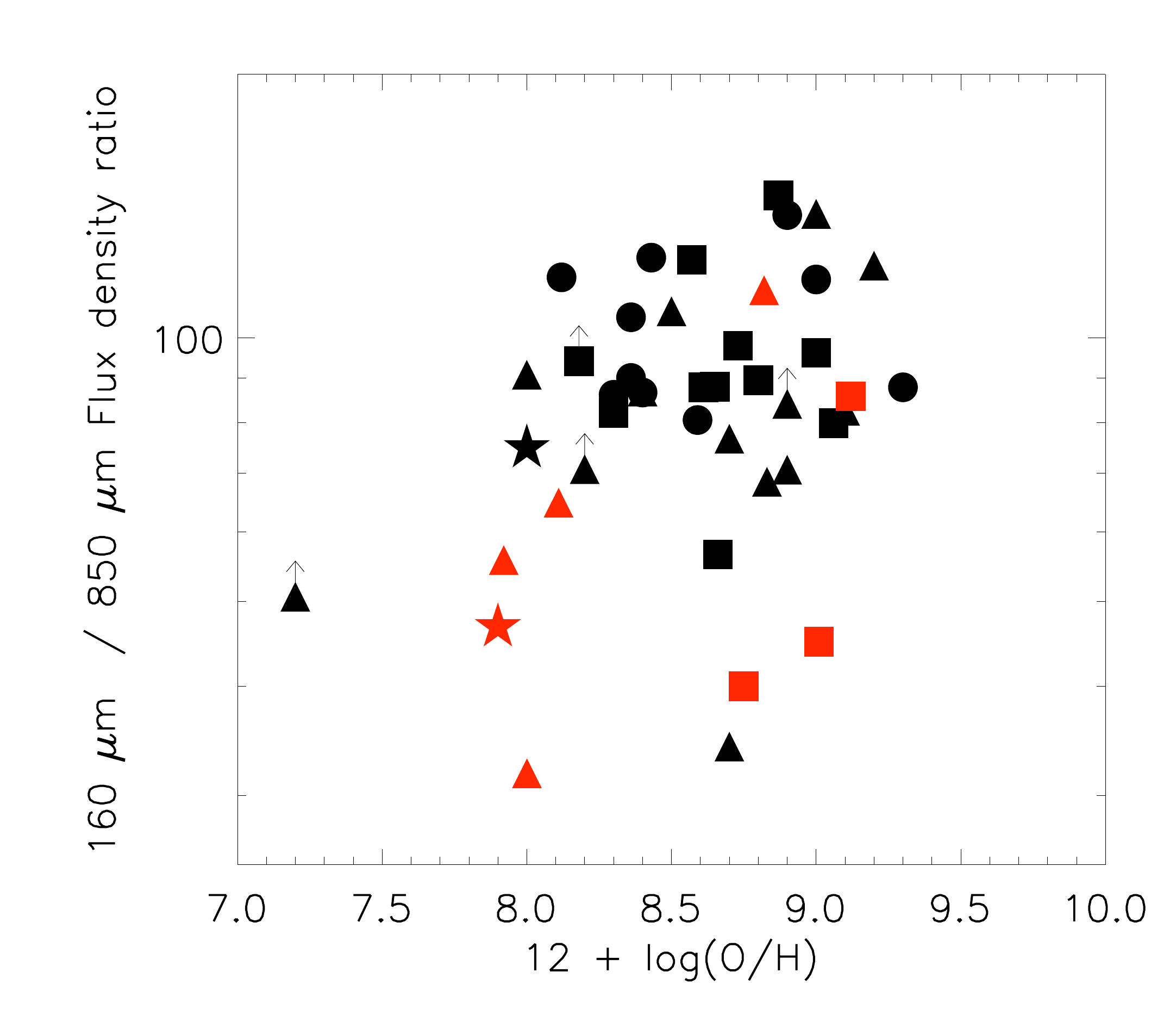} \\
         \end{tabular}
    \caption{  {\it a) }Correlation between the 160 \mic\ and the 850 \mic\ fluxes of the galaxies. A linear regression is overlaid (dashed line). When the galaxies do not possess an observational constraint at 850 \mic, the flux value was deduced from the SED model. {\it b) }160 \mic\ / 850 \mic\ flux density ratios as a function of metallicity. For convention on colors and symbols, see Fig.~\ref{Effects_of_the_submm_Figure}. Red symbols show the galaxies that present a submm excess and for which a cold dust (10K) component with an emissivity $\beta$=1 is added.}
    \label{160_850}
\end{figure}

We find that for 9 galaxies of our sample, a submm excess is detected (Fig.~\ref{SED_CD}). To study the physical conditions leading to an excess emission at submm wavelengths, we plot the 850 \mic\ flux as a function of the 160 \mic\ flux (Fig.~\ref{160_850}a). When the 160 \mic\ or 850 \mic\ fluxes are not directly observed, values are derived from the SED models we generated. Red symbols indicate galaxies where a submm excess is detected (SEDs modelled with an additional very cold dust component). While other galaxies follow a tight correlation, galaxies with submm excess show a 160-850 correlation that seems to be shifted. Figure~\ref{160_850}b shows the 160 \mic\ / 850 \mic\ flux density ratios as a function of metallicity. The 160 \mic\ to 850 \mic\ flux ratio is lower for galaxies in which a submm excess is detected (most of the time below $\sim$ 65, see the red symbols) but no clear correlation is found at this point with metallicity. Nevertheless, the sample is clearly lacking very low-metallicity galaxies (12+log(O/H)$<$7.9), which prevents us from studying the submm excess with an unbiaised metallicity coverage. 

We also note that a submm excess is detected in dustier galaxies (NGC~4303 and NGC~2903). In NGC~4303, the lack of far-IR measures between 100 and 450 \mic\ does not enable us to correctly sample the peak of the dust SED and the submm slope. Observations sampling this wavelength range (with \hersc\ for instance) would be required to confirm the excess in this galaxies. The dwarf galaxy He~2-10, which has a nearly solar metallicity, also shows a submm excess. We note that our fit has difficulties to fit the 100 and 160 \mic\ fluxes for this galaxy as well as for NGC~2903. However, the 850 \mic\ flux clearly exhibits an excess compared to the submm slope sampled by the 160 and 450 \mic\ fluxes.

Submm excess is most evidently detected in low-metallicity galaxies. If we assume that this excess is due to very cold dust, could it be possible that very cold dust is more apparent in dwarf galaxies compared to more metal rich galaxies? Indeed, on average, global SEDs of blue compact galaxies generally peak at relatively short wavelengths (30-60 \mic), compared to more quiescent, higher metallicity sources. For a metal-rich galaxy, peaking around 100 \mic\ for example, the submm emission at 870 \mic\ should be very difficult to distinguish from the emission of warm grains while in the case of a low-metallicity galaxy peaking shortward, the separation between two types of grain population temperatures, one warm and one very cold, may be easier to detect. We test this idea by modelling four galaxies with different metallicities (M51, M82, SBS 0335-052, Haro 11) with an independent very cold dust component, supposing that the mass and the properties ($\beta$, T) of the very cold dust (VCD) is the same, thus M$_{VCD}$/M$_{dust}$=constant. These tests lead to the inverse conclusion. The difference between the temperature of dust grains and the very cold grains is smaller in galaxies having higher metallicities. With a constant mass ratio, the emissivity ratio between the two components is thus higher in the dustier galaxies than in low-metallicity galaxies. If a very cold dust component exists with the same characteristics, our test here implies that it should be more evident as submm excess in dustier galaxies rather than in low-metallicity galaxies.  

\begin{figure}[h!]
   \centering
    \begin{tabular}{c }
      \includegraphics[width=8cm ,height=6.8cm]{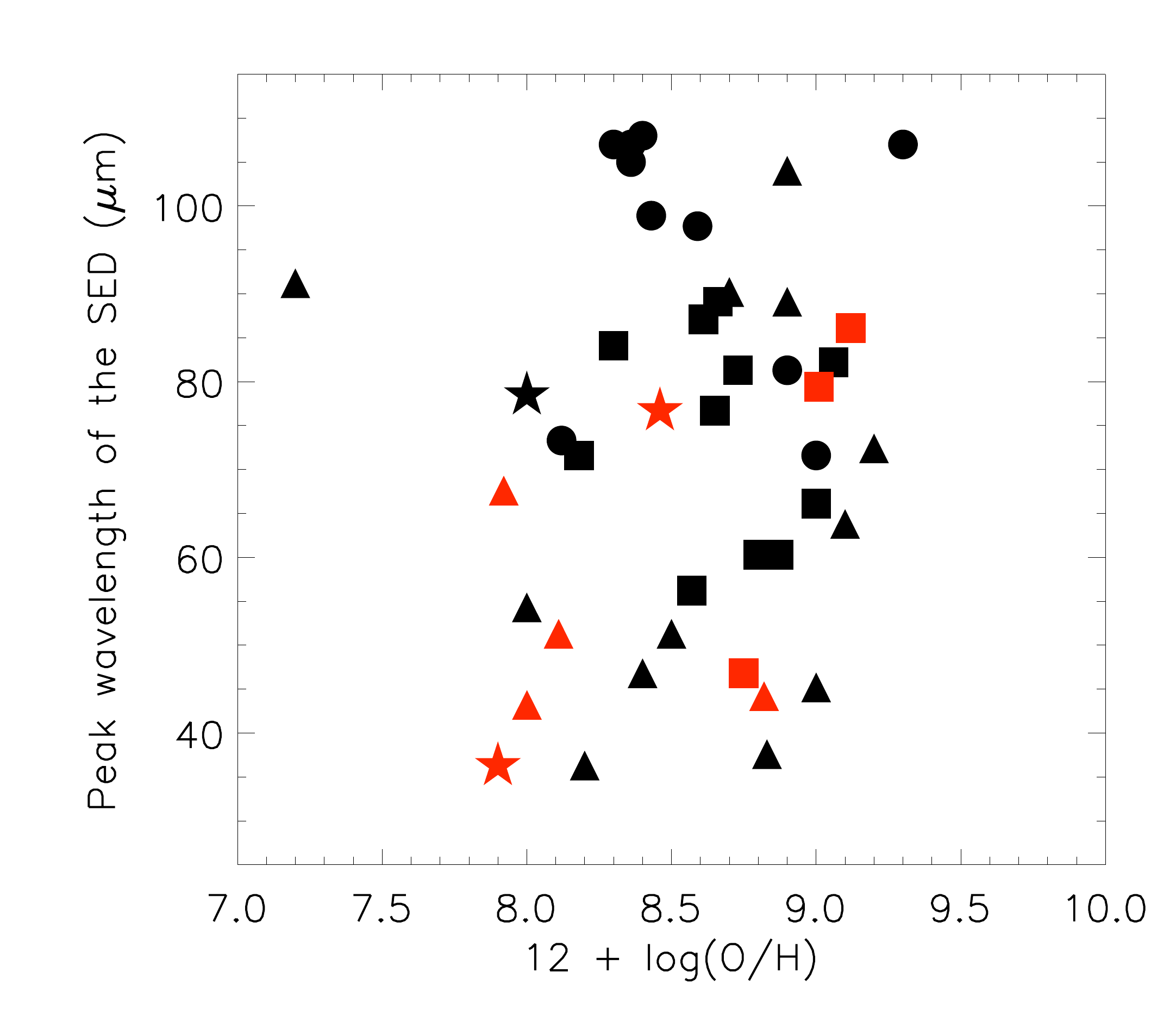} \\
         \end{tabular}
    \caption{ Wavelength of the peak of the dust emission as a function of  metallicity. For convention on colors and symbols, see Fig.~\ref{Effects_of_the_submm_Figure}. Red symbols show the galaxies that present a submm excess and for which a cold dust (10K) component with an emissivity $\beta$=1 is added.}
    \label{Peak_Z}
\end{figure}

The SEDs of low-metallicity galaxies usually show relatively ``warm" (f$_{70}$/f$_{160}$ $>$ 1) dust temperatures. For these ``warm" galaxies, the temperature of the SED is then well constrained but the flux at 870 \mic\ will significantly influence the total amount of dust while in ``colder" galaxies (f$_{70}$/f$_{160}$ $<$ 1), the SED model without submm data already leads to a large amount of dust. We plotted the ratio of the dust mass obtained with submm data over the dust mass without submm observations as a function of f$_{70}$/f$_{160}$ but did not find any correlation for our galaxies. 
We also plot the wavelength of the IR peak as a function of metallicity in Fig.~\ref{Peak_Z}. We do not clearly detect a correlation between the position of the dust emission peak and the metallicity of the galaxy. However, we note that a submm excess is not detected when the SED peaks above 85 \mic.
In some galaxies, the fact that a submm excess is not detected could also be due to an underestimate of the submm fluxes due to background subtraction issues (e.g. NGC~6946) or a lack of coverage of the complete emission in the submm observations (e.g. NGC~1097). Having several observational constraints in the submm slope of the SED would also help us to be more confident in the presence of submm excess: as shown in Fig.~\ref{SED_CD}, most of the galaxies where submm excess was detected were actually observed at different submm wavelengths (at 450 and 850 \mic). The addition of observational constraints will be a solution to avoid biases and sample a possible knee in the submm emission or a smoothness of the far-IR/submm slope, that could favor one of the two possible theories to explain the excess: a cold dust population or a change in the dust emissivity. 

{\revisedbis \hersc\ is currently observing in the 70 to 500 \mic\ wavelength range and will not only bring the lacking data to sample the SED but also the high resolution crucial to understand the origin of cold dust emission in galaxies. \hersc\ is also probing the coldest phases of dust in high-z galaxies. \citet{Santini2010} recently found that the dust content of their high-z submillimeter galaxies appears to be far higher than that expected from their metallicity, probably revealing different dust properties or dust growth mechanism in those environments. We caution the accuracy of dust masses estimates made without rest-frame submm observations. Accurate dust mass estimates in high-z galaxies are necessary to enable us to study the evolution of metals through cosmic time.  }


\section{Discussion and Conclusions}

We have gathered observational constraints of a large sample of galaxies from low-metallicity to metal-rich environments to study the influence of submm measurements on the total dust mass estimates and the consequences on the D/G. We use mid-IR and far-IR data from the literature. Our second sample is confined to galaxies having submm data beyond 160 \mic, namely 850 or 870 \mic, and sometimes 450 \mic, leading to a broad sample of 52 galaxies with submm observations. Submm observations probe the coldest dust grain emission of galaxies, a component that could account for a significant fraction of the total dust mass of galaxies. 

We model the SEDs of our sample and derive their total dust masses with and without submm fluxes to determine the importance of submm data in determining the dust SEDs. We find that submm data are crucial to accurately determine the dust masses. The error bars on the total dust mass estimates significantly decrease when using submm constraints. We show that submm data systematically affect the dust mass derived for our galaxies for different reasons: {\it a)} for dustier galaxies for which the SED usually peaks at longer wavelengths, submm fluxes are crucial to position the peak and the Rayleigh-Jeans slope of their SED and avoid an overestimation of the total dust mass of the galaxy usually observed when the SED modelling is performed using data at wavelengths only as long as 160 \mic; {\it b)} the submm wavelength domain sometimes harbours an excess that may imply a large amount of very cold dust. The lack of submm constraints will lead, in those cases, to an underestimate of the total dust mass of the galaxies. We caution the estimate of dust masses made without submm observations.

{\revisedbis We find that for 9 galaxies of our sample, mostly low-metallicity galaxies, our fiducial model can not fit the mid-IR 
to submm observational constraints and a submm excess is detected. We model this excess with a very cold dust component, using a temperature of 10K and a dust grain emissivity index of 1. The early results of \hersc\ also highlight a submm excess in dwarf galaxies \citep{OHalloran2010, Grossi2010}.
First results based on observations with the {\it Planck} mission in nearby galaxies have also lead to evidence of cold dust  \citep{Planck_collabo_2011_NearbyGalaxies}. In the Magellanic Clouds, \citet{Planck_collabo_2011_MagellanicClouds} have lead to the detection of an excess at submm/mm wavelengths. Average emissivity indexes $\beta$ of 1.5 and 1.2 were found for the LMC and SMC respectively, so flatter than observed in the Milky Way. Even taking the flatter slope into account, there is still evidence for a further flattening in the submm regime for the SMC, that has been modeled with a combination of thermal and spinning dust emission.} 

Gathering the gas mass information for our galaxies, we estimate the D/G of our sample. We find a tightened relation between the D/G and the metallicity of the galaxies when the models take into account submm observations. This effect is present in spite of our model assumption of cold dust for the submm excess. 

More submm observations of low-metallicity galaxies, {\revisedbis which will be obtained with \hersc\ and {\it Planck} will} extend the coverage to a broader range of metallicity and enable us to study how the D/G evolves in more metal poor environements. {\revisedbis The spatial resolution and sensitivity of \hersc\ combined with current ground based instruments (for example,  JCMT / SCUBA-2; APEX/Laboca) and ALMA will help us probe the submm emission of nearby galaxies in detail to better isolate the excess emission and understand the physics of the coldest phases of dust and the chemical enrichment thus drawing the global picture of the dust distribution within galaxies. While presently the nature of this submm excess seen in low metallicity environments is still not understood, the higher spatial resolution submm observations will help us disentangle the different scenarios: emission from very cold dust, possible changes in the dust grain emissivity, spinning dust etc. }




\bibliographystyle{apj}
\bibliography{mybiblio.bib}

\end{document}